\pdfoutput=1
\documentclass[a4paper,12pt]{article}
\usepackage{amsfonts}
\usepackage{mathrsfs}
\usepackage{amsmath}
\usepackage{amssymb}
\usepackage{framed}
\usepackage[medium]{titlesec}
\usepackage{bm}
\usepackage{cite}
\usepackage[normalem]{ulem}
\usepackage{extarrows}
\usepackage{slashed}
\usepackage{isodateo}
\usepackage{graphicx}
\usepackage{xcolor}
\usepackage[bookmarksnumbered=true,bookmarksopen=true]{hyperref}
 \hypersetup{colorlinks,%
             linkcolor=[rgb]{0,0.3,0.6}, %
             citecolor=[rgb]{0,0.3,0.6}, %
             urlcolor=[rgb]{0,0.3,0.6}}
\usepackage[hmargin=.7in,vmargin=1.1in]{geometry}
\usepackage{indentfirst}
\newcommand{\FR}[2]{\displaystyle\frac{\,{#1}\,}{#2}}

\newcommand{\n}{\nonumber}
\renewcommand{\rm}{\mathrm}

\graphicspath{{fig/}}

\def\bge{\begin{equation}}
\def\ede{\end{equation}}
\def\bga{\begin{aligned}}
\def\eda{\end{aligned}}
\def\bgp{\begin{pmatrix}}
\def\edp{\end{pmatrix}}
\def\bgs{\begin{subequations}}
\def\eds{\end{subequations}}
\newcommand{\order}[1]{\mathcal{O}({#1})}
\def\di{{\mathrm{d}}}

\def\mb{\mathbf}

\def\la{\langle}\def\ra{\rangle}

\def\to{\rightarrow}

\def\ii{\mathrm{i}}

\def\ga{\gamma}
\def\de{\delta}

\def\rh{\rho}
\def\si{\sigma}

\def\Mp{M_{\text{Pl}}}

\newcommand{\ob}[1]{\mkern 2mu \overline{\mkern -2mu #1 \mkern -2mu}\mkern 2mu}
\newcommand{\wt}[1]{\mkern 2mu \widetilde{\mkern -2mu #1 \mkern -2mu}\mkern 2mu}

\usepackage{enumerate}

\newcommand{\ti}{\wt t}

\makeatletter
\newcommand{\xRightarrow}[2][]{\ext@arrow 0359\Rightarrowfill@{#1}{#2}}
\makeatother

\newcommand{\beq}{\begin{eqnarray}}
\newcommand{\eeq}{\end{eqnarray}}
\newcommand{\beqa}{\begin{eqnarray}}
\newcommand{\eeqa}{\end{eqnarray}}

\def\OMIT#1{{}}
\newcommand{\lsim}{\mathrel{\rlap{\lower4pt\hbox{\hskip1pt$\sim$}}
    \raise1pt\hbox{$<$}}}         
\newcommand{\gsim}{\mathrel{\rlap{\lower4pt\hbox{\hskip1pt$\sim$}}
    \raise1pt\hbox{$>$}}}         

\begin{document} 

\title{\Large\textbf{A Cosmic Microscope for the Preheating Era}} 
\author{JiJi Fan$^a$\footnote{Email:\texttt{ jiji\char`_fan@brown.edu}}
~~~and~~~Zhong-Zhi Xianyu$^b$\footnote{Email:\texttt{ zxianyu@g.harvard.edu}}\\[2mm]
\normalsize{\emph{$^a$~Department of Physics, Brown University, Providence, RI 02912}}\\
\normalsize{\emph{$^b$~Department of Physics, Harvard University, 17 Oxford Street, Cambridge, MA 02138}}}

\date{}

\maketitle
 
\begin{abstract}
Light fields with spatially varying backgrounds can modulate cosmic preheating, and imprint the nonlinear effects of preheating dynamics at tiny scales on large scale fluctuations. This provides us a unique probe into the preheating era which we dub the ``cosmic microscope.'' We identify a distinctive effect of preheating on scalar perturbations that turns the Gaussian primordial fluctuations of a light scalar field into square waves, like a diode. The effect manifests itself as local non-Gaussianity. We present a model, ``modulated partial preheating," where this nonlinear effect is consistent with current observations and can be reached by near future cosmic probes. 

\end{abstract}

\section{Introduction} 

It is widely held that a period of cosmic inflation has happened in the early universe prior to the thermal big-bang phase, but it is nontrivial to connect inflation with the thermal big-bang. The transition is typically realized through reheating. Reheating is an abrupt and usually violent transition period, during which the inflaton energy is transferred to the thermal energy of other particles.\footnote{Notable exceptions include warm inflation, where inflation and the thermal big bang could be connected smoothly.} The thermal particles can be produced either through perturbative decays of the inflaton~\cite{Abbott:1982hn, Dolgov:1982th, Albrecht:1982mp}, or through various nonperturbative and out-of-equilibrium dynamics~\cite{Traschen:1990sw,Dolgov:1989us, Shtanov:1994ce, Kofman:1994rk, Boyanovsky:1995ud, Yoshimura:1995gc, Kaiser:1995fb, Kofman:1997yn}. The latter possibility is often called preheating, since it usually happens much faster and thus earlier than the perturbative decays. 

The preheating and reheating era dwells in a special corner of our knowledge about the cosmic history. On the one hand, it contains rich and complex dynamics that appeals to detailed study. On the other hand, the study of this era is plagued by a lack of observable effects. Preheating happens at scales tens of $e$-folds smaller than the CMB scale, and thus has little impact on large-scale observables. Only a handful of indirect observable effects of preheating are known, including a slight shift of inflation observables (scalar tilt and tensor-to-scalar ratio), and a stochastic gravitational wave background at high frequencies. A recent review could be found in~\cite{Lozanov:2019jxc}. It is thus desirable to look for more direct observable effects of preheating, especially the effects from its non-perturbative dynamics.

\begin{figure}[t]
\centering
\includegraphics[width=0.86\textwidth]{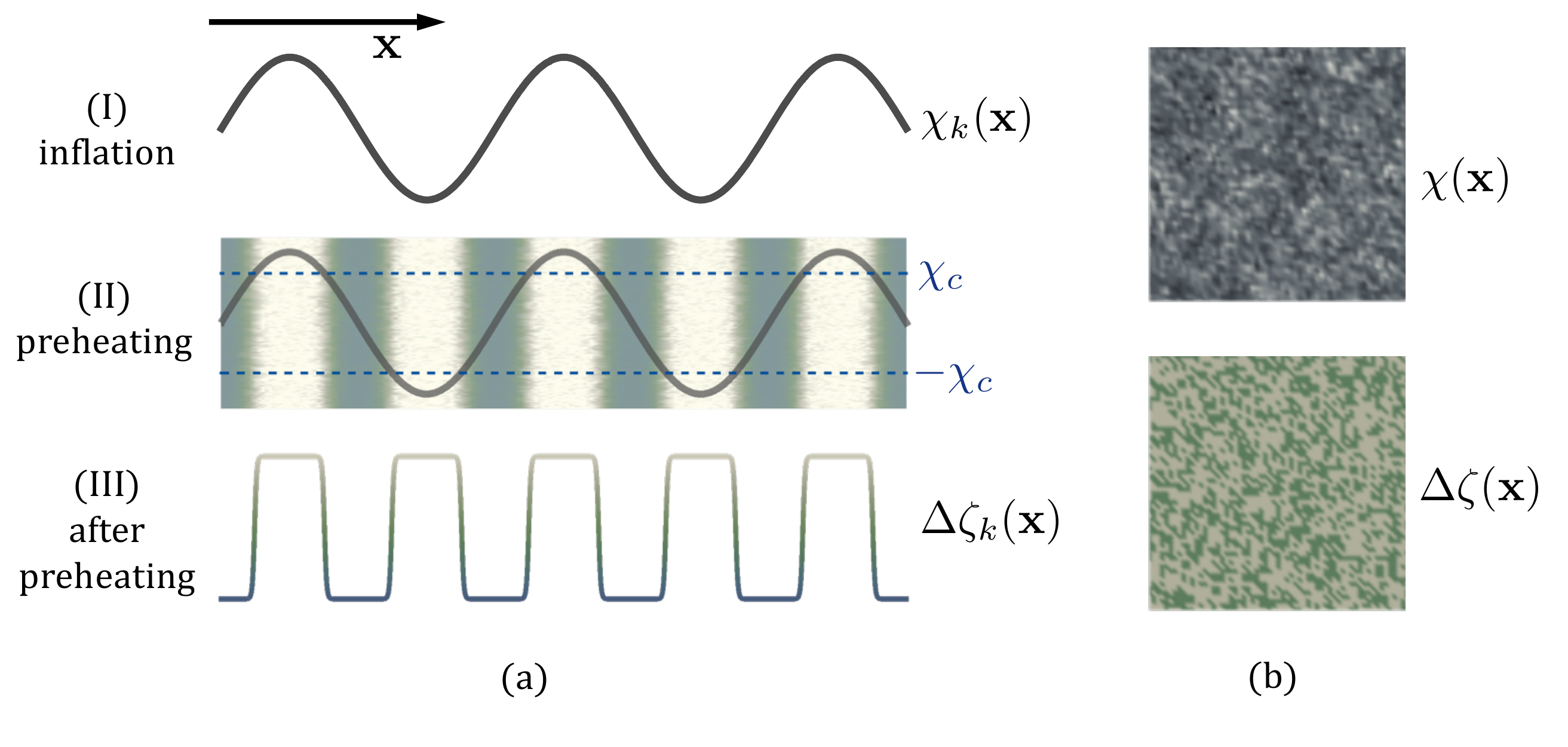} \vspace{-5mm}
\caption{Schematic picture of modulated preheating in our paper. Time flows from top to bottom. Column (a) is an illustration of the scenario with a single mode $k$. During inflation (I), a light scalar field $\chi$ acquires a spatially varying mode $\chi_k(\mb x)$, which modulates particle production during preheating (II). Preheating is triggered only when $|\chi|>\chi_c$ (light yellow region), and the region with $|\chi|<\chi_c$ (dark green region) has negligible preheating. This then contributes a ``square wave'' component to the curvature perturbation $\Delta\zeta_k(\mb x)$ (III). Column (b) shows a scale invariant Gaussian sample of $\chi(\mb x)$ and the corresponding $\Delta\zeta(\mb x)$ which is also scale invariant but almost bivalued.
}
\label{fig_modpreh}
\end{figure}

In this paper, we propose a scenario where the non-perturbative preheating dynamics could affect the large-scale fluctuation in a more direct manner. The key idea is to probe/perturb the preheating dynamics by an additional light scalar field $\chi$, which we call the \emph{modulating field} and which is ubiquitous in beyond SM physics.\footnote{The SM Higgs could be a candidate for the modulating field, though it is subject to more constraints. This includes the local non-Gaussianity generated by the Higgs self-interaction during the inflation, as well as the post-inflationary evolution of the Higgs. See \cite{Lu:2019tjj} for more discussions. } 
During inflation, $\chi$ acquires a nearly scale-invariant Gaussian background $\chi_0(\mb x)$. After inflation, $\chi_0(\mb x)$ becomes the effective coupling that controls preheating. Later in this paper we will show with an example that preheating is triggered only when $|\chi_0|$ is larger than a critical value $\chi_c$. Consequently the patches with $|\chi_0|$ below or above $\chi_c$ will experience very different expansion histories. In this way, the modulated preheating process will generate a characteristic curvature perturbation $\Delta\zeta(\mb x)$. We illustrate this process with a cartoon in Fig.\;\ref{fig_modpreh}.

In Fig.\;\ref{fig_modpreh}, we highlight the nonlinear nature of preheating, in contrast to modulated (perturbative) reheating. In the perturbative case, a small variation $\Delta\chi$ in the modulating field (and thus the coupling strength) would induce a small variation in the decay rate $\Delta\Gamma$ of the inflaton and therefore a small curvature perturbation, $\zeta=\Delta N$, where $N$ is the $e$-foldings of local expansion during the reheating era. In particular, $\Delta N$ depends smoothly on $\chi$ and a Taylor expansion in $\Delta \chi$ is always feasible, resulting in a \emph{linear} relation between the ``incoming'' $\chi$ wave and the ``outgoing'' $\zeta$ wave. However, in the modulated preheating case as we will see, the nonlinear nature of the particle production induces an abrupt dependence on $\chi$, effectively breaking regions with different $\chi_0$ into two phases. The resulting $\zeta$ will have an almost bivalued distribution. So the preheating dynamics has the effect of turning the original sinusoidal waves in $\chi$ into square waves in $\zeta$, like a nonlinear lens. 

One may wish to realize this scenario by requiring the $\chi$ field to modulate the non-perturbative inflaton decays. However, as we will show, such a scenario with all energy density $\rho_\text{total}$ going through preheating would result in a curvature perturbation $\zeta$ that is either nonlinear but too large, or of the right size but linear. To see the aforementioned nonlinear effect without invoking too large $\zeta$, we find it better to have only a small fraction of energy going through preheating, with the rest of energy released smoothly through perturbative reheating. In Sec.\;\ref{sec_model_pd}, we will describe a model realizing this partial preheating idea, in which we introduce a spectator field $\si$ in addition to the inflaton $\phi$. $\si$ has a subdominant but non-negligible energy density $\rho_\si$ during inflation. After inflation, the $\si$ sector will go through the modulated preheating while the inflaton still decays perturbatively. This model may appear a bit complicated with multiple fields but serves as a proof of principle. We will work out the corresponding curvature perturbation of this modulated partial preheating in Sec.\;\ref{sec:fractionalpreheating}. As one can expect, the curvature perturbation generated from this process is always suppressed by $\rh_\si/\rho_\text{total}$ and can be easily consistent with observation.  

It remains to see how to probe such a ``square wave'' effect. We explore this issue in Sec.\;\ref{sec_nG}. The main takeaway message is that the effect will manifest itself through the local non-Gaussianity of $n$-point correlators $(n\geq 3)$. Thanks to the modulating field and its nonlinear distortion, we could observe the nonlinear dynamics of preheating at super tiny scales directly at CMB scales. It is in this sense we call our scenario a {\it cosmic microscope.}

In general the ``square wave'' effect contributes only a small fraction of the curvature perturbation, while the major contribution is still from the inflaton fluctuation. Ideally, we may want to distill a bivalued distribution directly from the observed distribution of $\zeta$. But this is very challenging, since the modulated preheating happens within a Hubble patch, which is much smaller than the resolution, or the size of pixels, of any practical observation. Consequently, the observed value of $\zeta$ in each pixel is necessarily the average over many Hubble patches during preheating, making the result no longer bivalued. 
Therefore it is important to look for effects that survive this averaging process. We will show that the averaged $\zeta$ has little scale dependence, so one could not look for it in the power spectrum. However, there could be potentially large local non-Gaussianity. In particular, since we have a fixed square-wave relation between the incoming Gaussian $\chi_k$'s and outgoing $\zeta_k$'s, we would expect that the sizes of $n$-point correlations, for all $n\geq 3$, have a definite relation. At least in principle, if we could measure local non-Gaussianity in all $n$-point correlators, we should be able to identify this ``square wave'' component. But in practice, we will be able to access only a handful of them starting from $n=3$. Thus in this work we only focus on the 3-point function as a case study. 

We collect further discussions in Sec.\;\ref{sec_disc} and computational details in two appendices.

\paragraph{Comparison with previous works.} Readers familiar with the literature may recognize that some ingredients of our scenario also appear in scenarios of modulated reheating~\cite{Dvali:2003em, Kofman:2003nx, Suyama:2007bg, Ichikawa:2008ne, Lu:2019tjj} and modulated preheating~\cite{Kohri:2009ac}. In the former (latter) scenario, the inflaton perturbative decay (preheating) is controlled by the modulating field. In modulated reheating, the variation of the local $e$-fold number could be described by a {\it linear} function of $\chi$, as mentioned above. In the modulated preheating that has been studied in the literature, due to the observed size of $\zeta\sim 10^{-5}$, all the Hubble patches have to be either in an inefficient preheating phase with little energy transfer from inflaton to radiation~\cite{Kohri:2009ac} or in the efficient preheating phase~\cite{Enqvist:2012vx, Mazumdar:2015xka}. It is not allowed that some Hubble patches have little particle production and negligible back-reaction while others have sufficient particle production and back-reaction triggered by it. Staying within the same region without phase transitions makes it possible to generalize the idea of modulated reheating directly to preheating, but misses the interesting opportunity of probing the nonlinear effects of preheating.

In our scenario, the observational constraint is circumvented by requiring a spectator field with a sub-dominant energy fraction, instead of an inflaton, to undergo the nonlinear processes. To differentiate it from the modulated preheating in the literature, we will refer our mechanism as ``modulated partial preheating.'' The modulating field can scan different preheating phases in our scenario and probe the full nonlinear dynamics. The traditional view that preheating happens at too small a scale to be observable is circumvented by this scenario, which enhances the nonlinear effects of preheating all the way to CMB scales. Thus, again, we call this ``distorting the scale-invariant perturbation by the lens of preheating'' mechanism a cosmic microscope. 

Another class of preheating models that could generate large non-Gaussianities is the curvaton-type preheating scenario. The key difference is that in those scenarios, the additional scalar field with primordial fluctuations (the curvaton) directly participates in preheating either as a field created during preheating~\cite{Enqvist:2004ey, Enqvist:2005qu, Kohri:2009ac, Chambers:2007se, Bond:2009xx}, or the field responsible for the particle production~\cite{Chambers:2009ki}. All the couplings in the models are fixed. More details and references could be found in recent reviews of preheating~\cite{Amin:2014eta, Lozanov:2019jxc}.

\section{Overview of Modulated (P)reheating}
\label{sec_review}

In this section we review briefly (p)reheating and its modulated scenarios. This section is meant to provide a basic physical picture and some relevant formalism for non-experts. Readers who are familiar with the subject could safely skip this section. 

There is rich physics between the end of inflation and the onset of the thermal universe. As mentioned briefly in the introduction, there could be both perturbative decays and non-perturbative particle production. 
These will be reviewed in Sec.\;\ref{sec_preh_review}.

(P)reheating typically depends sensitively on various parameters. When such parameters vary over different spatial patches, (p)reheating, and therefore the expansion history of local universes, could be spatially dependent, too. This could generate additional curvature perturbations at large scales, providing us a unique window to probe the (p)reheating physics. The required spatially dependent parameter can come from the background value of a light field $\chi$. We will review this modulated scenario in Sec.\;\ref{sec_modreh_rev}.

\subsection{Reheating and Preheating}
\label{sec_preh_review}

In typical inflation scenarios, one needs a mechanism to stop inflation and then convert the inflaton energy to thermal radiation. The establishment of a thermal universe after inflation is called reheating. Typically, one can imagine that the inflaton falls into the bottom of its potential after inflation. Then it starts to oscillate and decay into other particles. Generically we expect the inflaton potential around the minimum to be quadratic, $V(\phi)\simeq \frac{1}{2}m_\phi^2\phi^2$. Then the equation governing the inflaton's evolution is 
\bge
  \ddot\phi(t)+3H\dot\phi(t)+m_\phi^2\phi(t)=0.
\ede
The Hubble parameter $H(t)$ decreases after the inflation, and the inflaton $\phi$ starts to oscillate when $H(t)<m_\phi$. A scalar oscillating in a quadratic potential behaves as cold matter when averaging over oscillation periods $1/m_\phi$, and has an effective equation of state $p=0$. So the universe expands as $a\propto t^{2/3}$, $H\simeq 2/(3t)$. Consequently, $\phi(t)\sim t^{-1}\cos(m_\phi t)$.

Here we assume that the perturbative decay rate $\Gamma$ of the inflaton is a constant and much smaller than $H(t)$ at the initial stage of inflaton oscillation, so that $\phi$ feels little friction from decays. Eventually, $H$ decreases to $\Gamma$ and the perturbative decays dominate the friction. The inflaton energy is completely transferred to radiation when $H(t)\simeq \Gamma$. At this point the universe is effectively dominated by radiation with an equation of state $p=\rho/3$ and thus expands as $a\propto t^{1/2}$.

The picture of perturbative decays described above is quite generic, but also often incomplete. There could be various sources of non-perturbative particle production, on top of the perturbative decays. Usually, such non-perturbative processes, called preheating, happen within a couple of $e$-folds after inflation while perturbative reheating usually happens significantly later. Another generic feature is that preheating could not complete the energy transfer from the mother scalar field to radiation so perturbative reheating is still needed to complete the transition after inflation. 

\begin{figure}[h]
\centering
\includegraphics[width=0.5\textwidth]{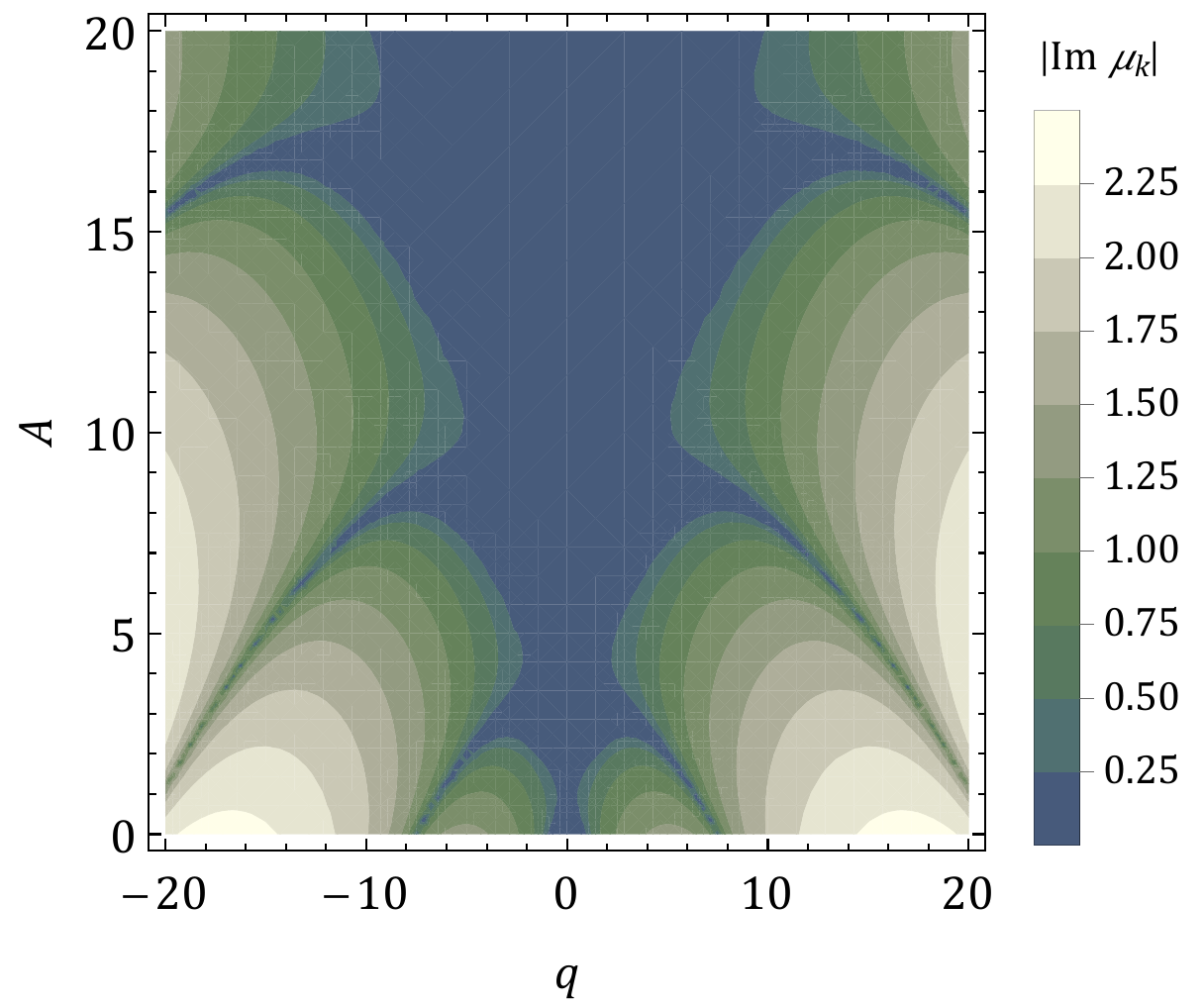} 
\caption{Floquet chart for Eq.~\eqref{eq:Mathieu}. The lighter regions correspond to unstable bands while the darker blue regions are stable with $\text{Im}\,\mu_k(A_k,q)=0$. }
\label{fig_floquet}
\end{figure}

Various non-perturbative mechanisms of particle production are known for preheating. See Ref.\cite{Lozanov:2019jxc} for a pedagogical introduction. The essential idea is that fast oscillations of the inflaton could introduce a time dependence in the effective mass of particles the inflaton couples to directly. In favorable parameter space, this time dependence can trigger proliferation of the associated particles through resonant production. Here we focus on one generic class of preheating mechanism known as tachyonic resonance~\cite{Dufaux:2006ee}. In the tachyonic resonance scenario, there is an interaction term, $-\frac{1}{2}g\phi\si^2$ between the inflaton $\phi$ and a scalar field $\si$. In Fourier space, the equation of motion for $\si(t,\mb k)$ at the linearized level (ignoring the self-interaction of $\si$) is,
\bge
  \ddot\si(t,\mb k)+3H(t)\dot\si(t,\mb k)+\big[m_\si^2+g\phi_0(t)+a^{-2}(t)k^2\big]\si(t,\mb k)=0.
\ede
To understand how $\si(t,\mb k)$ evolves in time, it is helpful to consider the same problem in a non-expanding space ($a=1$ and $H=0$), in which case the inflaton background is simply $\phi_0(t)=\ob\phi\cos(m_\phi t)$. Then the equation above reduces to the Mathieu equation,
\begin{align}
  \si''(z,\mb k)+\big[A_k+2q\cos(2z)\big]\si(z,\mb k)=0,
  \label{eq:Mathieu}
\end{align}
where $z\equiv m_\phi t/2$, $A_k\equiv 4(m_\si^2+k^2)/m_\phi^2$, $q\equiv 2g\ob\phi/m_\phi^2$, and the prime denotes $\di/\di z$. The most general solution of this equation is well known, and can be written as
\bge
\label{floq_sol}
  \si(z,\mb k)=e^{\ii \mu_k z}f^{(+)}_k(z)+e^{-\ii \mu_k z}f_k^{(-)}(z),
\ede
where $f^{(\pm)}$ are periodic functions, $f^{(\pm)}(z+\pi)=f^{(\pm)}(z)$ for any $z$, and the exponent $\mu_k=\mu_k(A_k,q)$. Therefore, in regions of parameter space with $\text{Im}\,\mu_k(A_k,q)\neq 0$, we will get an exponential enhancement of $\si$. The exponent $\mu_k(A_k,q)$ is conventionally displayed in the $(A_k,q)$ plane, known as the Floquet chart, as shown in Fig.\;\ref{fig_floquet}. 

The picture becomes more complicated after including the cosmic expansion and the back-reaction to the background geometry. When the expansion is slower than the characteristic time scale of the system, namely $H< m_\phi$ (which has to hold for the inflaton to oscillate), one could still use the adiabatic intuition. That is, one can think of $q=2g\bar\phi(t)/m_\phi^2$ as a time-dependent parameter and slowly decreasing towards zero as $\bar\phi(t)\propto 1/t$. (Note that physics is invariant under $q\to -q$.) Inspecting the Floquet chart, we see that the cosmic expansion would eventually bring the system out of resonance bands for a generic initial value of $q$, since the small $q$ region has no resonance. However, it is also possible that the expansion brings a previously non-resonant $q$ into a resonant band en route. This complicated process due to expansion of the universe is called stochastic resonance~\cite{Kofman:1997yn}. 
 In addition, once sufficiently many daughter particles are produced, they will back-react on the mother field and excite high-momentum modes. Both fields will eventually be fragmented spatially. Depending on the couplings in the model, preheating may end in the first stage with little particle production or go through the back-reaction stage, which we will refer to as \emph{efficient preheating} later. Note that efficient preheating in this paper does not imply full energy transfer from mother to daughter particles. It only means that an order one fraction of energy is transferred. We will discuss more about these interesting dynamics in our model in Sec.~\ref{sec_model_pd}.  

 For the purpose of this section, it suffices to point out that the equation of state of the universe can be altered quickly by preheating. Since the resonant production usually excites high $k$ modes through back-reaction, the equation of state $p=w\rho$ can quickly evolve towards $w=1/3$ in the case of efficient preheating.

\subsection{Modulated (P)reheating}
\label{sec_modreh_rev}

All the causal dynamics described above happens within a Hubble patch after inflation, whose size is extremely small compared with the CMB scale. In the simplest picture of single field inflation, different patches evolve similarly. In particular, preheating would happen either in all patches or in none. 

There are alternatives when we go beyond the single field picture. Suppose that, in addition to the inflaton $\phi$, there is a light scalar field $\chi$ with mass $m_\chi\ll H_\text{inf}$, the Hubble scale during inflation. During inflation, it develops a spatially varying background $\chi=\chi(\mb x)$ at cosmological scales. The fluctuation $\de\chi$ is almost scale invariant since $\chi$ is light, and typically has the size $\de\chi\sim H_\text{inf}$. Then, after inflation, the $\chi$ background within each Hubble patch will be a constant over space (but not necessarily in time\footnote{Depending on its potential $V(\chi)$, the $\chi$ field would in general have nontrivial time dependence after inflation. For instance, if $V(\chi)=\frac{1}{2}m_\chi^2\chi^2$, then $\chi$ will start oscillating when $H$ drops to $m_\chi$. In this section we will ignore the evolution of $\chi$ for simplicity, without affecting the main conclusion. Practically, one can achieve this by assuming $m_\chi\ll H(t)$ for all $t$ of interest. However, the post-inflationary evolution could be crucial in some cases, see for example \cite{Lu:2019tjj}.}). 

It is plausible that the variation in the background value $\chi_0$ would affect the expansion history in each individual Hubble patch. Consider perturbative reheating first. In this case, the cubic coupling $\frac{1}{2}g\phi\si^2$ that has been considered in the previous section is actually from a dim-4 operator $\frac{1}{2}g'\chi\phi\si^2$ with $g=g'\chi_0$. Then the perturbative decay rate $\Gamma(\phi\to\si\si)\propto \chi_0^2$. As a result, the time of reheating $t_\text{reh}$, determined by $\Gamma=H(t_\text{reh})$, depends on $\chi_0$ through its dependence on $\Gamma$. Consequently, varying $\chi$ over large scales induces a change of expansion histories across different Hubble patches, and thus provides a new source generating curvature perturbations. This is known as the modulated reheating scenario, and we call $\chi$ the modulating field from now on. 

The idea above can be neatly formulated by the so-called $\de N$ formalism~\cite{Starobinsky:1986fxa, Sasaki:1995aw, Lyth:2004gb}. In short, the curvature perturbation $\zeta(t_2,\mb x)$ at a later time $t_2$, long after the reheating completes, receives contributions from two distinct sources. One is the ``primordial perturbation'' generated during inflation $\zeta(t_1,\mb x)$, where $t_1$ is chosen before inflation ends but after the modes of interest go outside of the horizon. The other one is the contribution from the ``reheating'' era, namely between $t_1$ and $t_2$, in the form of perturbed $e$-folding number $N$,
\bge
  \zeta(t_2,\mb x)=\zeta(t_1,\mb x)+ \de N(t_1,t_2,\mb x), ~~~~~\de N(t_1,t_2,\mb x)\equiv N(t_1,t_2,\mb x)-\ob{N}(t_1,t_2,\mb x),
\ede  
where $\ob{N}$ is the unperturbed $e$-folding number.  
In the single field story, the curvature perturbation comes mainly from $\zeta(t_1,\mb x)$ and $\de N$ can be neglected. But in the modulated reheating, $\de N$ is non-negligible. In fact, assuming the equation of state is $w=0$ before reheating and $w=1/3$ afterwards, we can find,
\bge
  N(t_1,t_2,\mb x)=\int_{t_1}^{t_\text{reh}}\di t  H(t)+\int_{t_\text{reh}}^{t_2}\di t H(t)=\FR{2}{3}\log\FR{t_\text{reh}}{t_1}+\FR{1}{2}\log\FR{t_2}{t_\text{reh}}.
\ede
Here we have used $H=2/(3t)$ for matter domination and $H=1/(2t)$ for radiation domination.
Now we perturb $\Gamma$, and thereby perturb $t_\text{reh}$. From the relation $H=2/(3t)$ $(t\leq t_\text{reh})$ and $H(t_\text{reh})=\Gamma$, we get $\de t_\text{reh}/t_\text{reh}=-\de\Gamma/\Gamma$, and therefore,
\bge
  \de N=-\FR{1}{6}\FR{\de\Gamma}{\Gamma}.
\ede
Since $\de\Gamma\propto\de\chi$, the curvature perturbation gets a contribution from $\de\chi$ in this scenario.

The formalism above can be directly adapted to include preheating. Instead of the perturbative decay rate $\Gamma$, the preheating efficiency is controlled by the $\mu_k$ parameter introduced in Eq.~\eqref{floq_sol}, which depends on the coupling $g$, and thus on the modulating field $\chi$, in a highly nonlinear way, as illustrated in Fig.\;\ref{fig_floquet}. The parameter $\mu_k$ then determines the time preheating occurs, $t_\text{pre}$, which roughly sets the boundary between effective matter domination (due to inflaton oscillation) and effective radiation domination. A variation in $t_\text{pre}$ would perturb the expansion history, similar to the perturbative case discussed above. The crucial difference, though, is that the parameter $\mu_k$ (and thus $t_\text{pre}$) depends on $\chi$ in a highly nonlinear way. Consequently, a small linear change in $\chi$ could induce a large nonlinear perturbation in the expansion history. We would then expect a dramatic nonlinear contribution to the curvature perturbation through this mechanism. In fact, we will show in the following sections that the curvature perturbation induced by such a modulated preheating scenario is almost always too large to be consistent with the observed value $\zeta\sim 10^{-5}$. This leads us to consider a preheating scenario with sufficient nonlinear effects in the curvature perturbation but a small overall amplitude of $\zeta$. The model will be discussed in the following sections.

\section{The Model and Phase Diagram}
\label{sec_model_pd}

In this section we will first present our main model in which a fraction of curvature perturbation is from modulated preheating. In this model, 1) the size of the curvature fluctuation can be easily made consistent with the observation while 2) it still carries the non-perturbative information of the preheating processes. We argue in last section that these two properties are difficult to be realized simultaneously in the inflaton preheating scenario. To circumvent this problem, we introduce a spectator field $\si$ which goes through the non-perturbative decay, while the inflaton still decays perturbatively.

To be concrete, we will still use the trilinear coupling, analogous to what we have reviewed in the previous section, but between the spectator $\si$ and an additional scalar $\psi$, and the associated tachyonic resonance mechanism. We will also present a phase diagram of the mechanism in the parameter space of the model.

\subsection{The Model}
We consider a model containing four scalar fields: the inflaton $\phi$; a spectator field $\si$ with a sub-dominant but not negligible energy fraction during inflation; the daughter scalar particle $\psi$ that could be produced through non-adiabatic and nonlinear processes; and finally, the modulating field $\chi$ with a spatially varying background that dictates the coupling between $\si$ and $\psi$. The full scalar potential is given by
\beq
\label{Vpotential}
V = V(\phi) + V(\chi)+ V(\sigma) + \frac{g(\chi)}{2} \si \psi^2 + \frac{\lambda}{4} \psi^4,
\eeq
where we assume that the sector of $\si$ and $\psi$ is decoupled from the inflaton sector. The potential is subject to the following constraints: 
\begin{enumerate}
\item The potential is bounded from below. This requires 
\beq
\label{pertcond}
\frac{1}{2\lambda}\left(\frac{g(\chi)}{m_\si}\right)^2 \leq 1,
\eeq
where $m_\si\equiv V''(\si_\text{min})$. 

\item $\si$'s energy density is sub-dominant during inflation, $\rho_\sigma \sim V(\si_0) \ll H_\text{inf}^2 \Mp^2$ with $\si_0$ the background value of $\si$ and $H_\text{inf}$ the Hubble parameter during inflation. $\Mp \simeq 2.4 \times 10^{18}$ GeV is the reduced Planck mass. Here we assume that $V(\si_0)$ also has the form of a slow-roll potential. That is, $V(\si)$ is flat enough during the inflation and the kinetic energy $\frac{1}{2}\dot\si_0^2$ can be neglected.

\item $m_\si > H_\text{end}$ immediately at the end of inflation so that $\si$ field starts oscillating in its potential around the minimum $\si_\text{min}$ at the end of the inflation. Combining with Constraint 2, we see that it is favorable to have a hill-top like potential for $\si$.

\item For simplicity we will assume that $V(\chi)$ is very flat so that $\chi$, once it has acquired some fluctuations during inflation, remains roughly constant during the whole period of preheating. This is not necessary and one can consider a more complicated situation with a non-trivial post-inflationary evolution of $\chi$. 

\end{enumerate}
The fraction of energy density in the $\si$ field at the end of inflation is 
\beq
\frac{\rho_\si}{\rho_{\rm {tot}}} \sim \frac{m_\si^2 \si_0^2}{H_\text{end}^2 \Mp^2} = \left(\frac{m_\si \si_0}{H_\text{end}^2}\right)^2 \left(\frac{H_\text{end}}{\Mp}\right)^2 \sim 10^{-10} \left(\frac{m_\si \si_0}{H_\text{end}^2}\right)^2 \left(\frac{H_{\rm{end}}}{10^{13}\,\rm{GeV}}\right)^2. 
\eeq
Since the highest inflationary Hubble scale is bounded to be $10^{13}$ GeV~\cite{Akrami:2018odb} at least in simple models, to have the fraction larger than $10^{-10}$, we need to have 
\beq
\si_0 \gg H_{\rm{end}},
\eeq
assuming $m_\si \gtrsim H_{\rm{end}}$ but not much larger than $H_{\rm{end}}$.

\subsection{Tachyonic Resonance of the Spectator Field} 
\label{sec:phasediagram}

\subsubsection{Linearized Equation of Motion}

Now we study the production of $\psi$ particles. For this purpose we will first consider the linearized equation of motion for $\psi$, ignoring the back-reaction. 

We assume that inflaton oscillates after inflation in a quadratic potential before its perturbative reheating so that it generates a matter domination background with a scale factor
\beq
a(t) = a_0 \left(\frac{t}{t_0}\right)^{2/3} =  \left(\frac{m_\si t}{2}\right)^{2/3},
\eeq
where in the second step, we set the initial time and scale factor to be $t_0 = 2/m_\si$ and $a_0 = 1$ for convenience of calculations latter. The coherent oscillation of the $\si$ field also redshifts as matter with an amplitude
\bge
\si (t) = \frac{\si_0}{a(t)^{3/2}} \cos \left[m_\si (t- t_0)\right].
\ede
Given the $\si$ background above, together with the potential in Eq.~\eqref{Vpotential}, we can find the \emph{linearized} equation of motion for $\psi$. In Fourier space, the equation for a comoving mode $k$ is,
\bge
 \ddot{\psi}_k + 3 H \dot{\psi}_k + \left(\frac{k^2}{a^2} + g\si(t)  \right) \psi_k = 0.
\ede
Here we treat $g$ as a constant in each Hubble patch since the horizon size after inflation is much smaller than the wavelength of $\chi$ fluctuations and the value of $\chi$ is roughly a constant in each patch. Upon substitution 
\beq
\wt t=t/t_0=m_\si t/2, \quad X\equiv a^{3/2}\psi, \quad A_k^0\equiv(2k/m_\si)^2, \quad \rm{and} \quad q_0=2g\si_0/m_\si^2, 
\eeq
the equation above transforms to a form of the Mathieu equation as reviewed before, except for the redshift dependences:
\bge
 \frac{d^2 X_k}{d\wt t^{\;2}} +\left(\frac{A_k^0}{\wt t^{4/3}} +  \frac{2 q_0}{\wt t} \cos \left[2 (\wt t- 1)\right]\right) X_k =0. 
 \label{eq:eom}
\ede
We have reviewed the solution to this type of equation without cosmic expansion in Sec.~\ref{sec_preh_review}. Below we will include the expansion effects.

The most important consequence of expansion of the universe is that it will eventually kill the particle production process when $A_k^0/\wt{t}^{4/3}\ll 1$ and $4 q_0/\wt{t} \ll 1$. Thus the {\it comoving} occupation number $n_k$ of mode $k$ is expected to grow exponentially first and then reach a plateau. Indeed this could be confirmed both analytically and numerically. It is found analytically in Ref.~\cite{Abolhasani:2009nb} that in the exponentially growing stage, for large enough $q_0 > 5$, the comoving occupation number as a function of time could be approximated by 
\beq
\ln n_k \simeq \frac{8 \alpha}{\pi} \sqrt{q_0 \; \wt t}  - \frac{12 \alpha A_k^0}{\pi \sqrt{q_0}} \; \wt t^{\; 1/6} + \cdots,
\label{eq:lnnk}
\eeq
where $\alpha \simeq 0.85$ and $\cdots$ contain constant and phase terms which either don't depend on time or depend weakly on time. According to Eq.~\eqref{eq:lnnk}, there are three important features of the growth: 
\begin{enumerate}
\item Larger $q_0$ (or equivalently, larger trilinear coupling) leads to faster exponential growth.

\item The cosmic expansion slows down the growth so that the comoving occupation number $n_k$ grows with an exponent proportional to $\sqrt{t}$ instead of $t$ as in the case of a non-expanding universe. 

\item Increasing the wave number $k$ would reduce the growth exponent. For fixed $t$, the largest wave number $k_\text{max}$ that can be excited can be found by asking the two terms in Eq.~\eqref{eq:lnnk} to cancel. This leads to 
\beq
k_{\rm {max}} \simeq \frac{m_\si}{2} \sqrt{q_0} \, \wt t^{\;1/6}.
\label{eq:kmax}
\eeq
\end{enumerate}

\begin{figure}[h]
\centering
\includegraphics[width=\textwidth]{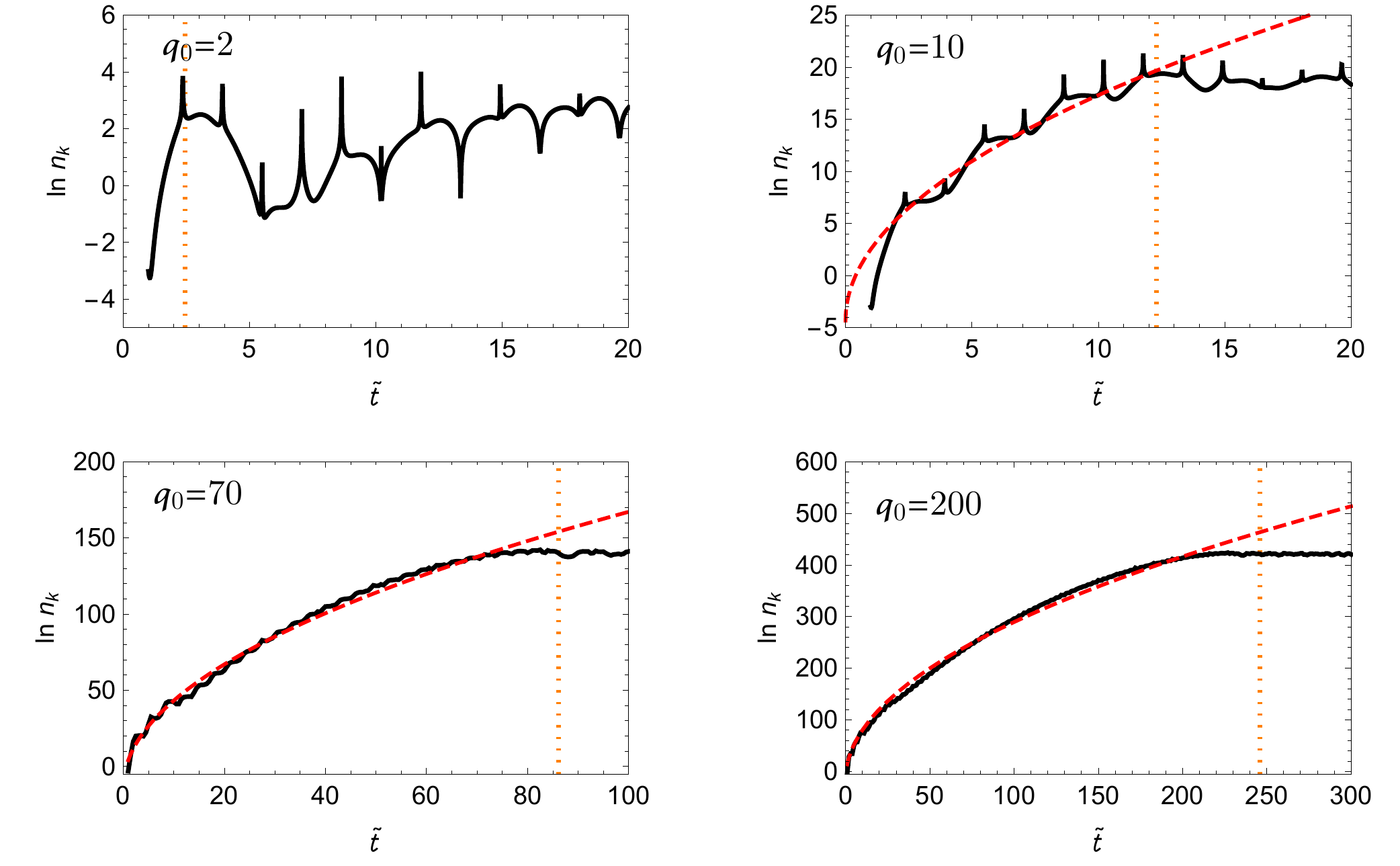} 
\caption{The logarithm of the comoving occupation number $n_k$ as a function of time for $k, A_k^0 = 0$ and different choices of $q_0$. Numerical results are shown by black curves. The red dashed lines show the analytical estimates based on Eq.~\eqref{eq:lnnk}. The orange dotted lines indicate $\wt t_\text{term}$ estimated using Eq.~\eqref{eq:tend}.  }
\label{fig:lnnk}
\end{figure}

In the linearized analysis, particle production is terminated by the cosmic expansion at $t_{\rm {term}}$, which satisfies~\cite{Abolhasani:2009nb}
\beq
\frac{q_0}{\wt t_{\rm {term}}} \simeq 0.81 \left(1-\frac{A_k^0}{\wt t_{\rm {term}}^{\; 4/3}}\right), \quad \rm{where} \quad \wt t_{\rm {term}} = \frac{m_\si t_{\rm {term}}}{2}. 
\label{eq:tend}
\eeq
The analytical formula in Eqs.\eqref{eq:lnnk} and \eqref{eq:tend} agree well with numerical results obtained by solving Eq.~\eqref{eq:eom} numerically, as shown in Fig.~\ref{fig:lnnk}. Note that Eq.~\eqref{eq:lnnk} breaks down when $q_0 \lesssim 5$. For smaller $q_0$, there is simply not much particle production as demonstrated in the first panel of Fig.~\ref{fig:lnnk}.

\subsubsection{Phase Diagram}
Particle production transfers energy from $\si$ to $\psi$. In the linearized analysis, the fraction of energy density of $\psi$ at any time before $t_{\rm {term}}$ could be estimated to be 
\beq
\delta(t) &\equiv & \frac{\rho_\psi(t)}{\rho_{\si+\psi}(t)} = \frac{2\rho_\psi(t) a(t)^3}{m_\si^2 \si_0^2} \simeq e^{\frac{8 \alpha}{\pi} \sqrt{q_0 \ti}} \frac{q_0} {4608 \alpha^2 \ti} \left(\frac{m_\si}{\si_0}\right)^2.
\label{eq:delta}
\eeq
More details of the derivation for the equation above could be found in Appendix~\ref{app:fraction}. Requiring $\delta(t_{\rm {term}}(k=0)) = 0.1$, we find that 
\beq
q_0=q_c \simeq 30  + 0.84 \log \left(\frac{\phi_0/m_\phi}{10^{13}}\right). 
\label{eq:qc}
\eeq
When $q_0 <q_c$ ($q_0 > q_c$), the fraction of energy density in $\chi$ is below (above) $10\%$ at $t_{\rm {term}}$. 

If at a time, $t_{\rm {com}}$, there is a significant fraction (e.g. 10\%) of energy density in the $\psi$ field comparable to what remains in the $\si$ field, the back-reaction from $\psi$ to the mother field $\si$ through the trilinear coupling could not be ignored. The linearized analysis breaks down. Instead a full numerical simulation based on coupled equations of motion is needed as in Ref.~\cite{Amin:2019qrx}. Once the back-reaction is effective, the system will quickly evolve into an energy equipartition state with similar amounts of energy in the $\psi$ and $\si$ fields. The effective equation of state, 
\beq
w_{\rm {sub}}= p_{\rm {sub}}/\rho_{\rm {sub}}
\eeq 
of the sub-system of $\si$ and $\psi$, also quickly rises to a plateau $\sim 0.3$, signaling a mixed radiation-matter state~\cite{Amin:2019qrx}. In the discussions below, we will approximate the asymptotic value of $w_{\rm{sub}}$ to be that of radiation $1/3$, for simplicity of the calculation without modifying the conclusions. 

One additional complication arises from the self-interaction of the $\psi$ field. If the quartic coupling $\lambda$ is large enough and sufficiently many $\psi$ particles are produced, the self-interaction of $\psi$ turns into an effective positive mass term $\lambda \langle \psi^2 \rangle \psi^2$ and slows down the particle production. 

\begin{figure}[h!]
\centering
\includegraphics[width=0.9\textwidth]{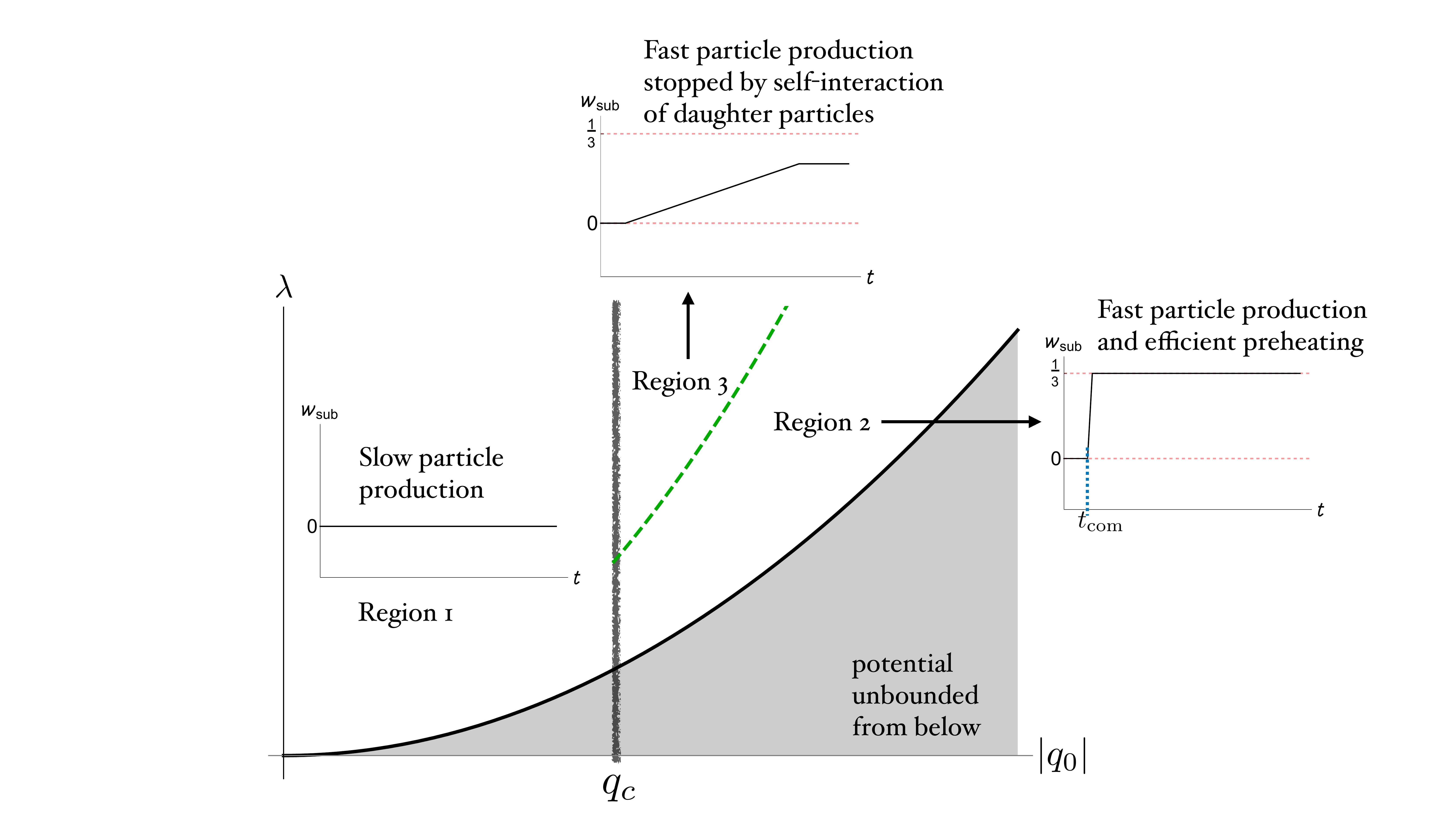} 
\caption{Schematic phase diagram of the spectator tachyonic resonance mechanism in the parameter space $(q_0, \lambda)$. $w_{\rm {sub}} = p_{\rm {sub}}/\rho_{\rm {sub}}$ defined for the (sub-)system of $\si$ and $\psi$. The black curve corresponds to $\lambda =\frac{q_0^2}{8} \left(\frac{m_\si}{\si_0}\right)^2$, below which the potential is unbounded. The green dashed curve corresponds to a large enough $\lambda$, which leads to a large positive $\lambda \langle \psi^2 \rangle \psi^2$ to stop particle production and separates region 2 and 3.  }
\label{fig:schematic}
\end{figure}

Putting the considerations together, we have three possible ``phases" (regions) in the $(q_0, \lambda)$ plane, as depicted in Fig.~\ref{fig:schematic}:
\begin{itemize}
\item Region 1: $q_0 < q_c$. In this region, $t_{\rm {term}} < t_{\rm {com}}$. Particle production is stopped by expansion of the universe before there are comparable energies in $\si$ and $\psi$. Back-reaction is negligible and the analysis based on the linearized equation of motion is valid. The fraction of energy transferred from $\si$ to $\psi$ is {\it exponentially} sensitive to $q_0$, as shown in Eq.~\eqref{eq:delta}, but it is always $\ll 1$. Thus $w_{\rm{sub}}$ is always approximately zero. 
\item Region 2: $q_0 > q_c$ and $ 0.1 \lesssim \frac{1}{2\lambda} \left(\frac{g}{m_\si}\right)^2 \leq 1 $. In this regime, $t_{\rm {term}} > t_{\rm {com}}$.  At around $t_{\rm {com}}$, the linearized equation of motion breaks down and the system will quickly rises to a constant $w_{\rm{sub}}$ close to 1/3. When $q_0$ is large enough, this is approximately a step function in time. Notice that there could be more details in the time evolution of $w_{\rm{sub}}$ as shown in Fig. 2 in Ref.~\cite{Amin:2019qrx}. For instance, before settling down to the asymptotic value, $w_{\rm{sub}}$ could have an intermediate oscillation stage. 
\item Region 3: $q_0 > q_c$ and $\frac{1}{2\lambda} \left(\frac{g}{m_\si}\right)^2 \gtrsim 0.1 $. In this region, back-reaction is non-negligible as in region 2 and the linearized analysis breaks down. In addition, the quartic interaction, $\lambda \langle \psi^2 \rangle \psi^2$ becomes important and comparable to the trilinear term $g \langle \si \rangle \psi^2$. It acts as an effective positive mass term for $\psi$ and slows down the tachyonic particle production process. This is probably the most tricky region with no simple description of the time evolution. The schematic picture of time evolution of $w_{\rm{sub}}$ in Fig.~\ref{fig:schematic} merely serves as an illustration.
\end{itemize}

Note that numerical simulations in Ref.~\cite{Amin:2019qrx} supporting this phase diagram are implemented in the inflaton preheating scenario. We have carried out lattice simulations for the preheating of the spectator field $\si$ with a fixed cosmic expansion and checked that this phase diagram still holds. More details of the simulation results could be found in Appendix~\ref{app:lattice}.

\section{Modulated Partial Preheating} 
\label{sec:fractionalpreheating}

At the end of last section, we presented the ``phase diagram'' for our preheating scenario, as in Fig.\;\ref{fig:schematic}. Throughout our treatment in the previous section, we took the coupling $g$ responsible for preheating to be a constant. It is indeed a constant in time within each Hubble patch if the modulating field $\chi$ has little evolution during preheating, which we assume for simplicity. But across many different Hubble patches, the coupling $g(\chi)$ will vary \emph{spatially}, due to the spatial fluctuations of $\chi$ developed during inflation. Therefore, by looking at different Hubble patches, we effectively scan different regions in the phase diagram. In this section, we will figure out the resulting density perturbation $\zeta$, using the $\de N$ formalism reviewed in Sec.~\ref{sec_modreh_rev}.

There exist several possible scenarios for which region(s) in Fig.~\ref{fig:schematic} the modulating field $\chi$ scans. The simplest ones are that the modulating field only scans one region, e.g., region 1, as discussed in the original modulated inflaton preheating paper~\cite{Kohri:2009ac} or region 3, as discussed in refs~\cite{Enqvist:2012vx, Mazumdar:2015xka}. In fact, these are the only possibilities allowed for modulated inflaton preheating given the constraint on the size of the curvature fluctuation. The more interesting possibility, which only appears when we consider modulated partial preheating for a spectator field, is that the variation of the modulating field covers a broad range of $q_0 = 2 g \sigma_0/m_\sigma^2$ and thus scans multiple regions. For instance, the modulating field scans Regions 1 and 2. As the value of the modulating field varies from one Hubble patch to the other, some patches have negligible particle production with relatively simple dynamics that could be described by linearized equations of motion (Region 1) while the other patches have efficient preheating with complicated dynamics due to back-reaction (Region 2). This possibility could be realized if the trilinear coupling depends on the light modulating field as 
\beq
g(\chi) = \frac{\chi^2}{\Lambda}, 
\eeq
where the cutoff $\Lambda \geq \si_0$ for the effective field theory to be valid. Since the modulating field is light during inflation, its fluctuation $\delta \chi \sim H_{\rm{inf}}$ while the mean value $\langle \chi_0 \rangle $ could be around or above $H_{\rm{inf}}$ for $\chi$ with a very flat potential. Then we have the mean value of $q_0$ and the variation in $q_0$ to be 
\beq
\langle q_0  \rangle &\sim & \frac{\langle \chi_0 \rangle^2 \sigma_0}{\Lambda m_\sigma^2} \lesssim \frac{\langle \chi_0 \rangle^2}{m_\sigma^2} \lesssim \frac{\langle \chi_0 \rangle^2}{H_{\rm{inf}}^2},  \nonumber \\
\delta q_0 & \sim & \frac{\langle \chi_0 \rangle (\delta \chi) \sigma_0}{\Lambda m_\sigma^2} \lesssim \frac{\langle \chi_0 \rangle (\delta \chi)}{m_\sigma^2}\lesssim \frac{\langle \chi_0 \rangle(\delta \chi)}{H_{\rm{inf}}^2},
\eeq
where we use that $m_\sigma \gtrsim H_{\rm{end}}$, the Hubble scale at the end of inflation, and assume that Hubble scales do not change much throughout inflation $H_{\rm{end}} \sim H_{\rm{inf}}$.  
To scan different regions, $\langle q_0  \rangle$ has to be close to $q_c$, which is of order (10 -100) as shown in Eq.~\eqref{eq:qc}. Thus, we need 
\beq
\langle \chi_0 \rangle \sim {\cal O} (\text{a few} - 10) H_{\text{inf}},
\eeq
and $\delta q_0 \sim {\cal O} (0.1) \langle q_0  \rangle$.

In this case, one could model the time evolution of $w_{\rm {sub}}$ in different regions simply as
\beq
\rm{Region} \; 1: \quad w_{\rm {sub}} &\equiv & \frac{p_{\rm {sub}}}{\rho_{\rm {sub}}} \simeq  0,   \\
\rm{Region} \; 2: \quad w_{\rm {sub}} & \simeq &  0, \quad \quad \; \;  \; \ti_0< \ti< \ti_{\rm {com}}(q_0(\chi)), \nonumber \\
&\simeq & 1/3  \quad \quad \;  \ti> \ti_{\rm {com}}(q_0(\chi))
\label{eq:wsub}
\eeq
where $\ti_{\rm {com}} = t_{\rm{com}}m_\si/2$ is the rescaled time when there are comparable energies in $\si$ and $\psi$. 
Now we could see in a more quantitative way beyond the argument we sketch at the end of Sec.~\ref{sec_modreh_rev} why we require the spectator field instead of the inflaton to preheat. If $\si$ is the inflaton, $w_{\rm{sub}}$ is then the equation of state for the entire system. Varying $q_0$ to jump from Region 1 to Region 2 leads to $\order{1}$ differences in $w$ between different Hubble patches. The number of $e$-folds, $N \propto \int H dt =  \frac{2}{3(1+w)} \int dt/t$, will change by $\order{1}$, too. The variation of $N$ yields an ${\cal O}(1)$ contribution to the curvature fluctuation $\zeta$, which is too large to be consistent with the observation $\zeta \sim 10^{-5}$.  
Therefore, we consider the initial energy density of $\si$ to be a small fraction, $\gamma$, of the total energy density shortly after inflation at time $t_0$:
\beq
\gamma \equiv \frac{\rho_{\si}(t_0)}{\rho_{\rm {tot}}(t_0)}\ll 1. 
\eeq

\begin{figure}[h!]
\centering
\includegraphics[width=0.8\textwidth]{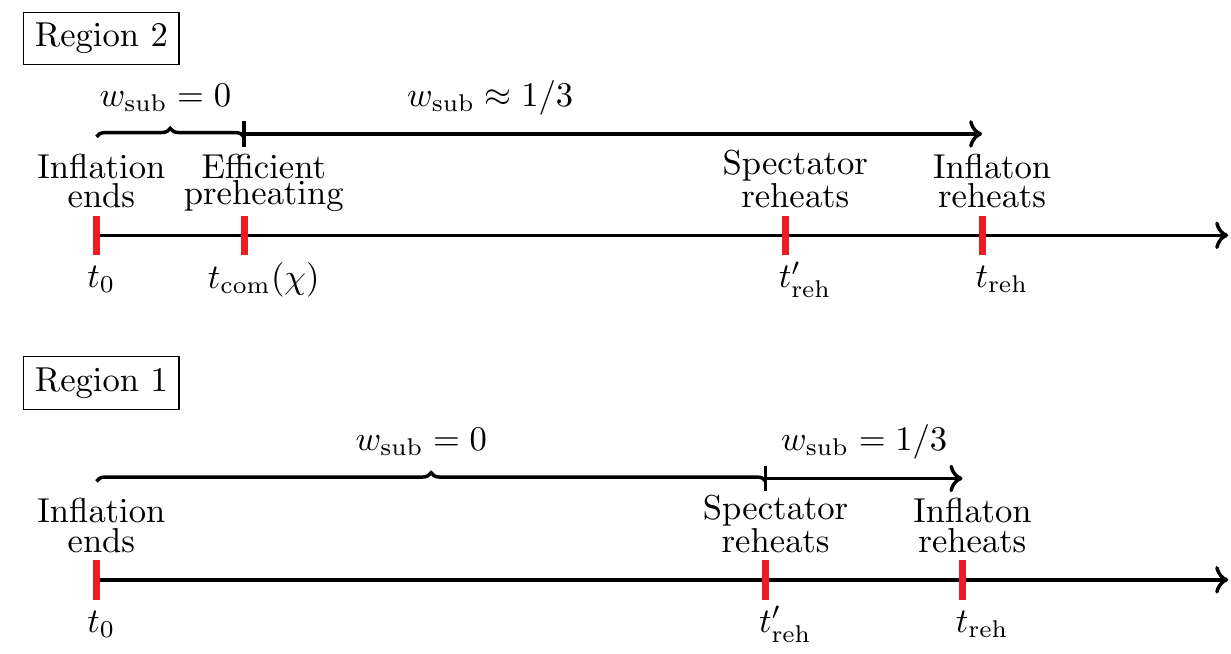} 
\caption{Characteristic time scales in Region 2 with efficient preheating of $\sigma$ and Region 1 with negligible preheating of $\sigma$. We also indicate $w_{\rm {sub}}$. }
\label{fig:timearrows}
\end{figure}

We focus on the scenario where the modulating field scans region 1 and region 2. After $t_0$, there are three relevant time scales in the scenario: $t_{\text {com}}$, when the efficient preheating happens in region 2; $t_{\rm{reh}}^\prime$, the perturbative reheating time of the spectator field $\si$; and $t_{\rm{reh}}$, the perturbative reheating time of the inflaton, present in both regions. We assume that $\si$ reheats before the inflaton so that $t_{\rm{reh}}^\prime < t_{\rm{reh}}$. Though the trilinear coupling $\si \psi^2$ could also lead to perturbative reheating via $\si \to \psi \psi$, we assume that $\si$ reheats through other channels, e.g., decays to two fermions. This assumption is to make reheating and preheating of $\si$ independent of each other to simplify the discussions. We hold the decay widths of $\si$ and the inflaton, $\Gamma_\si$ and $\Gamma_\phi$, to be spatially invariant so that the total energy of the universe when $\phi$ ($\si$) reheats is a constant across all Hubble patches. Note that there is a subtlety here: due to different expansion histories, $t_{\rm{reh}}$ and $t_{\rm{reh}}^\prime$ vary slightly in different patches though the total energy densities at these times are invariant. 
The relevant time scales and the flow of events in both regions are depicted in Fig.~\ref{fig:timearrows}. 

In the $\de N$ formalism, we compute the number of $e$-folds between two uniform energy density slices (which are not necessarily constant time slices) in each Hubble patch. In our scenario, we choose the two slices to be a slice shortly after inflation at $t_0$ and the inflaton reheating slice at $t_{\rm{reh}}$. 
Since the inflaton $\phi$ redshifts as matter, $N$ between the end of inflation and inflaton reheating is given by 
\beq
N = \frac{1}{3} \log \frac{\rho_\phi(t_0)}{\rho_\phi(t_{\rm {reh}})},
\eeq
where $\rho_{\phi}(t_0)$ is the energy density of inflaton immediately after inflation and $\rho_\phi(t_{\rm {reh}})$ is the energy density of inflaton when it reheats at $t_{\rm {reh}}$. Since the inflaton reheats when the Hubble scale drops to be its perturbative decay width $\Gamma_\phi$ (which is taken to be fixed), we have 
\beq
\rho_{\rm {tot}}(t_{\rm {reh}}) = 3\Mp^2 H(t_{\rm {reh}})^2 = 3  \Mp^2 \Gamma_\phi^2= \rho_\phi(t_{\rm {reh}})+ \rho_{\si}(t_{\rm {reh}}),
\eeq
where $\rho_{\si}(t_{\text{reh}})$ is the energy density of the decay products from either preheating or reheating of $\si$ at the inflaton reheating time.

For patches in region 2 with $|q_0| > q_c$, approximating $\si$ and its daughters to redshift as radiation after $t_{\rm {com}}$ as in Eq.~\eqref{eq:wsub},\footnote{Strictly speaking, efficient preheating only transfers an order one amount of energy from $\si$ to $\psi$ and the leftover $\si$ still needs to reheat to turn into radiation fully. Yet in the tachyonic resonance scenario, $w_{\rm{sub}}$ reaches a plateau close to $1/3$ after preheating and we ignore the small difference between $w_{\rm{sub}}$ and 1/3, which won't affect the conclusions.}  we have 
\beq
\rho_\phi (t_{\text{reh}}) =\rho_{\rm {tot}}(t_{\text{reh}}) - \rho_{\si} (t_{\text{reh}})=  \rho_{\rm {tot}} (t_{\text{reh}}) \left[1 - \gamma \left(\frac{a(t_{\rm {com}})}{a(t_{\text{reh}})}\right)\right].
\eeq
In the equation above, we take into account that after $\si$'s preheating, the energy density of $\si$ and its decay products dilutes faster than that of the inflaton and thus the fraction of energy density stored in the sub-system is reduced further by a factor of $\frac{a(t_{\rm {com}})}{a(t_{\text{reh}})}$. 
Given that to the leading order of $\gamma$, the whole system redshifts as matter, we have
\beq
&&\frac{a(t_{\rm {com}})}{a(t_{\text{reh}})}  \simeq \left(\frac{t_{\rm {com}}}{t_{\text{reh}}}\right)^{2/3} \simeq \left(\frac{H(t_{\text{reh}})}{H(t_{\rm {com}})}\right)^{2/3} \simeq \left(\frac{3}{2} \, \Gamma_\phi  \, t_{\rm {com}}\right)^{2/3},
\eeq
where we use that in a matter domination phase, $a(t) \propto t^{2/3}$, $H= \frac{2}{3t}$ and $H_{\rm {reh}} = \Gamma_\phi$.
Combining all the equations above, we obtain that in patches with efficient spectator preheating (Region 2), 
\beq
N_{2} &=& \frac{1}{3} \log \frac{\rho_\phi(t_0)}{\rho_{\rm {tot}}(t_{\text{reh}})} - \frac{1}{3} \log \left[1 - \gamma \left(\frac{3\Gamma_\phi}{2} t_{\rm {com}}\right)^{2/3}\right],  \nonumber \\
&\simeq & \frac{1}{3} \log \frac{\rho_\phi(t_0)}{\rho_{\rm {tot}}(t_{\text{reh}})} + \frac{\gamma}{3} \left(\frac{3\Gamma_\phi}{2} t_{\rm {com}}\right)^{2/3}.
  \eeq

In patches without efficient preheating (region 1) with $|q_0| < q_c$, we have
\beq
N_1 &= & \frac{1}{3} \log \frac{\rho_\phi(t_0)}{\rho_\phi(t_{\text{reh}})} = \frac{1}{3} \log \frac{\rho_\phi(t_0)}{\rho_{\rm {tot}}(t_{\text{reh}}) } -\frac{1}{3} \log \left[1- \gamma \frac{a(t_\text{reh}')}{a(t_{\text{reh}})}\right] \nonumber \\
&\simeq & \frac{1}{3} \log \frac{\rho_\phi(t_0)}{\rho_{\rm {tot}}(t_{\text{reh}})} + \frac{\gamma}{3} \left(\frac{\Gamma_\phi}{\Gamma_\si}\right)^{2/3}.
\eeq
Note that the formula above is based on our assumption that $\si$ reheats before inflaton reheats. If the order is switched, after the inflaton reheats, $\si$ still redshifts as matter and its energy fraction increases. We will not discuss this possibility further. 

In summary, the $e$-folding number $N$ as a function of $q_0$ (and thus the modulating field $\chi$) is 
\beq
N (q_0)  = N_2 + A \; \vartheta\big(q_c-|q_0|\big),
\label{eq:efolds}
\eeq
where $\vartheta$ is the Heaveside step function and 
\beq
A=N_1- N_2 \simeq \frac{\gamma}{3} \left(\frac{\Gamma_\phi}{\Gamma_\si}\right)^{2/3} - \frac{\gamma}{3} \left(\frac{3\Gamma_\phi}{2} t_{\rm {com}}\right)^{2/3}  \simeq \frac{\gamma}{3} \left(\frac{\Gamma_\phi}{\Gamma_\si}\right)^{2/3} , 
\label{eq:diff}
\eeq
where in the last step, we use the fact that preheating happens (much) earlier before reheating, $t_{\rm{com}} \ll 1/\Gamma_\si < 1/\Gamma_\phi$. Thus the value of $A$ is independent of the the modulating field $\chi$. 
So the local $e$-folding number across different Hubble patches behaves as a square wave as the trilinear coupling varies, as shown in panel a(III) of Fig.~\ref{fig_modpreh}. Note that $A$ is proportional to $\gamma$, the initial energy fraction of $\si$, but suppressed further by $\left(\Gamma_\phi/\Gamma_\si\right)^{2/3}$. In the limit that inflaton reheating happens much later than the reheating of $\si$, $A \to 0$.

\section{Non-Gaussianity as a Probe of Preheating History}
\label{sec_nG}

In the last section we show that the modulated partial preheating is able to generate a contribution to the curvature perturbation in a highly nonlinear fashion but with a suppressed amplitude. The total curvature fluctuation $\zeta$ has two contributions,
\begin{align}
  \zeta=\zeta_\text{inf}+\zeta_\text{mp},
\end{align}
where $\zeta_\text{inf}=-(H_{\rm{inf}}/\dot\phi_0)\de\phi$ is from the usual inflaton fluctuation, and $\zeta_\text{mp}$ is from the modulated preheating,
\begin{align}
  \zeta_\text{mp}(\mb x)=\delta N(\mb x)=A\vartheta\Big(\chi_c-|\chi_0(\mb x)|\Big),
\label{zetadN}
\end{align}
where $A \simeq  \gamma/3 \left(\Gamma_\phi/\Gamma_\si \right)^{2/3}$, $\chi_c=\sqrt{q_c\Lambda m_\si^2/(2\si_0)}$, and $\chi_0(\mb x)$ is the background of the modulating field $\chi$ generated during inflation, which we take to be Gaussian and scale invariant. We assume that the mixing between the inflaton and $\chi$ fluctuations is small and thus ignore cross correlators.

In Fig.\;\ref{fig_modpreh} we demonstrate with a simple cartoon that $\zeta_\text{mp}(\mb x)$ looks like a square wave for a single $k$-mode of $\chi$. But the scale-invariant modulating field $\chi$ is a superposition of many $k$-modes and there is no simple square-wave like behavior after the superposition. Thus we need to look elsewhere for its observable consequences. 

In this section we will show that one possible set of observables is local non-Gaussianities of $\zeta$. The relation in Eq.~\eqref{zetadN} distorts the original Gaussian spectrum of $\chi$ in a distinctive way, and it is in principle possible to reconstruct this ``square-wave function'' in Eq.~\eqref{zetadN} by measuring all $n$-point correlations of $\zeta$. Practically this is surely very challenging, and we have access to only a handful of $n$-point correlations that could possibly be extracted from data, starting from $n=3$. So in this section we will use the 3-point function as an example and leave a more exhaustive study for future work. We will also assume that the non-Gaussianity is dominated by the contributions from $\zeta_\text{mp}$, while the contribution from $\zeta_\text{inf}$ is negligible. 

Since we work in a fully nonlinear regime where the usual perturbative expansion does not work, we will first present a formalism in Sec.\;\ref{sec_ng_gf} that works for this special case. A similar treatment was developed to compute non-Gaussianity in curvaton-type preheating in Ref.~\cite{Suyama:2013dqa}. In Sec.\;\ref{sec_ng_mpp} we will apply this formalism to the study of modulated partial preheating. 

Since we will focus only on $\zeta_\text{mp}$, we will drop the subscript in the rest of this section unless confusion may arise. 

\subsection{General Formalism}
\label{sec_ng_gf}


The nonlinear nature of the preheating process means that $\zeta$ depends on the background value of the modulating field $\chi$ in a nontrivial way, invalidating the Taylor expansion. So we must treat the function $\zeta=\zeta(\chi)$ nonlinearly. This will introduce characteristic local-type non-Gaussianities of $\zeta$ at large scales. Since the functional dependence $\zeta(\mb x)=\zeta(\chi(\mb x))$ is local in position space, it is more convenient to consider the fluctuations in position space instead of momentum space. Going from the usual formulation in momentum space to position space introduces some subtleties, which we will discuss first before presenting the general formalism. 

The vacuum fluctuation of the light modulating field $\chi$ ($m_\chi \ll H_{\rm{inf}}$) is governed by a classical Gaussian distribution outside the horizon. In momentum space, it has a zero mean $\la\de \chi_{\mb k}\ra=0$. The 2-point function (power spectrum) can be parameterized as $\la\de \chi_{\mb k_1}\de \chi_{\mb k_2}\ra= (2\pi)^3\de^{(3)}(\mb k_1+\mb k_2)P_\chi(k_1)$.
For a light field $\chi$, the power spectrum will be nearly scale-invariant, 
\beq
P_\chi(k)\simeq  \frac{2\pi^2\Delta_\chi^2}{k^3}.
\eeq
We would like to understand how this looks like in position space. To this end one could try a straightforward Fourier transform as follows,
\begin{align}
  \la\de \chi (\mb x_1)\de \chi(\mb x_2)\ra=\int\FR{\di^3\mb k_1}{(2\pi)^3}\FR{\di^3\mb k_2}{(2\pi)^3}e^{\ii(\mb k_1\cdot\mb x_1+\mb k_2\cdot\mb x_2)}\la\de \chi_{\mb k_1}\de \chi_{\mb k_2}\ra=\int\FR{\di^3\mb k}{(2\pi)^3}e^{\ii\mb k\cdot(\mb x-\mb y)}P_\chi(k),
\end{align}
which is divergent. The divergence is in the IR, meaning that we would see increasing correlations by going to larger scales. But observation-wise we are always limited by the total amount of data available to us -- the size of the observable universe, which introduces an IR cutoff $L$, explicitly breaking the scale invariance. As a result, the position-space correlator becomes 
\bge
 \la\de \chi(\mb x_1)\de \chi(\mb x_2)\ra_L=\Delta_\chi^2\bigg[\text{sinc}\bigg(\FR{|\mb x_1-\mb x_2|}{L}\bigg)-\text{Ci}\bigg(\FR{|\mb x_1-\mb x_2|}{L}\bigg)\bigg],
\ede
where $\text{sinc}(x) = \sin(x)/x$ and the cosine integral $\text{Ci}(x) = \int_x^\infty \cos(t)/t  \; \di t$.  
For $|\mb x_1-\mb x_2|\ll L$, we have,
\begin{align}
\label{chi2pt}
  \la\de \chi(\mb x_1)\de \chi(\mb x_2)\ra_L\simeq \Delta_\chi^2\log\FR{\ob L}{|\mb x_1-\mb x_2|},
\end{align}
where $\ob L\equiv L e^{1-\ga_E}$ with $\ga_E$ the Euler gamma constant. 
The 2-point function also diverges in the coincident point limit. This divergence is cut off by the resolution of the sampling. Suppose the size of the pixel is $\ell$, the above 2-point function tells us that $\chi(\mb x)$ at a given point is a random Gaussian variable with a zero mean and a variance 
\beq
\la \de \chi^2(\mb x)\ra \simeq \Delta_\chi^2\log\left(\frac{\ob L}{\ell}\right).
\label{eq:coincidence}
\eeq
 
As a check of the position space correlator, we could Fourier transform it back to the momentum space and obtain
\begin{align}
  \int\di^3x\, e^{-\ii\mb k\cdot\mb x}\Delta_\chi^2\log\FR{\ob L}{|\mb x|}\simeq &~\FR{4\pi}{k^3}\Delta_\chi^2\big[\text{Si}(kL)-\sin(kL)\big],\n\\
  \simeq &~\FR{2\pi^2}{k^3}\Delta_\chi^2\Big[1-\FR{2}{\pi}\sin(kL)+\mathcal{O}(\FR{1}{kL})\Big], 
\end{align}
where the sine integral $\text{Si}(x) = \int_0^x \sin(t)/t\; \di t$ and in the second line, we keep leading terms in the $kL \gg 1$ limit. 
Indeed we recover the original scale invariant power spectrum in momentum space, with corrections of the form $\sin (kL)$. This is a typical artifact of the sharp cutoff. Such fast oscillatory terms have characteristic scale $L$, which sets the resolution of measuring $k$ and is averaged to zero.

Now we consider the non-Gaussianities induced by the nonlinear preheating processes. By nonlinear we mean that the curvature perturbation $\zeta$ generated during preheating is a nonlinear function of the modulating field fluctuation, $\zeta(\mb x)= \zeta(\de \chi(\mb x))$. In the standard $\delta N$ formalism, one often Taylor expands this function in terms of small fluctuation $\de \chi$ as $\zeta(\mb x) \supset \de N = N - \overline{N}=N_a\de \chi(\mb x)+N_b (\de \chi(\mb x))^2+\cdots$, which directly leads to a nonzero local non-Gaussiainity even when $\de \chi$ itself is purely Gaussian. However, as we have shown, in modulated partial preheating, $\de N$ is not a smooth function and a Taylor expansion is impossible, which makes a non-perturbative treatment essential. The question is then to find the correlation function $\la\zeta(\de \chi)\cdots\zeta(\de \chi)\ra$ given the function $\zeta=\zeta(\de \chi)$ and the fact that $\de \chi$ is scale-invariant and Gaussian.  As we mentioned above, it is easier to compute in position space as the relation $\zeta=\zeta(\chi)$ is local in position. 
 
In general, the $n$-point correlation function could be computed as a functional integral
\begin{align}
  \Big\la \mathcal{O}_1[\chi(\mb x_1)]\cdots \mathcal{O}_n[\chi(\mb x_n)]\Big\ra=&~\mathcal{N}\int\bigg[\prod_{\mb x}\di \chi(\mb x)\bigg]\exp\bigg[-\FR{1}{2}\int\di^3x\di^3y \; \chi(\mb x)\mathcal{D}(\mb x,\mb y)\chi(\mb y)\bigg]\n\\
  &\times \mathcal{O}_1[\chi(\mb x_1)]\cdots \mathcal{O}_n[\chi(\mb x_n)],
\end{align}
where $\mathcal{N}$ is the overall normalization factor and $\mathcal{D}(\mb x,\mb y)$ is the functional inverse of the 2-point correlator $\mathcal{G}(\mb x,\mb y)\equiv\la \chi(\mb x) \chi(\mb y)\ra$, satisfying
\bge
  \int\di^3\mb y\,\mathcal{D}(\mb x,\mb y)\mathcal{G}(\mb y,\mb z)=\de(\mb x-\mb z).
\ede 
To calculate the correlator of curvature perturbations $\zeta$, we can just take $\mathcal{O}_i[\chi(\mb x_i)]=\zeta(\chi(\mb x_i))$ and carry out the functional integral above. However, there is a simpler way that turns the functional integral above into an ordinary integral~\cite{Suyama:2013dqa}. The idea is to expand the function $\zeta(\chi)$ in plane waves $e^{\ii\omega \chi}$ rather than in powers $\chi^n$. The plane wave $e^{\ii\omega \chi}$ can be treated as a generating function of moments $\la \chi^n\ra$. To see how it works, let's define
\bge
\label{zetatilde}
  \zeta(\chi)=\int\FR{\di\omega}{2\pi}e^{\ii\omega \chi}\wt\zeta_\omega,
\ede
which is point-wise in $\mb x$. Then, the $n$-point correlator of the curvature fluctuation $\zeta$ can be found by
\bge
\label{zetan_ft}
  \la\zeta(\mb x_1)\cdots\zeta(\mb x_n)\ra=\int\FR{\di\omega_1}{2\pi}\cdots\FR{\di\omega_n}{2\pi}\,\wt\zeta_{\omega_1}\cdots\wt\zeta_{\omega_n}\Big\la e^{\ii\omega_1\chi(\mb x_1)}\cdots e^{\ii\omega_n \chi(\mb x_n)}\Big\ra.
\ede
The correlator on the right side could be calculated explicitly,
\begin{align}
  \Big\la e^{\ii\int\di^3\mb x\,\omega(\mb x)\chi(\mb x)}\Big\ra=&~\mathcal{N}\int\bigg[\prod_{\mb x}\di\chi(\mb x)\bigg]e^{-\frac{1}{2}\int\di^3x\di^3y \, \chi(\mb x)\mathcal{D}(\mb x,\mb y)\chi(\mb y) +\ii\int\di^3\mb x\,\omega(\mb x)\chi(\mb x)}\n\\
  =&\exp\bigg[-\FR{1}{2}\int\di^3\mb x\di^3\mb y\,\omega(\mb x)\omega(\mb y)\la \chi(\mb x) \chi(\mb y)\ra\bigg].
\end{align}
To calculate the correlator in Eq.~\eqref{zetan_ft}, we can simply take $\omega(\mb x)=\sum_{i=1}^n\omega_i\de(\mb x-\mb x_i)$, and get
\begin{align}
\la\zeta(\mb x_1)\cdots\zeta(\mb x_n)\ra=\int\FR{\di\omega_1}{2\pi}\cdots\FR{\di\omega_n}{2\pi}\,\wt\zeta_{\omega_1}\cdots\wt\zeta_{\omega_n}\exp\bigg[-\FR{1}{2}\sum_{i,j=1}^n\omega_i\omega_j\la \chi(\mb x_i) \chi(\mb x_j)\ra\bigg].
\label{eq:generalformalism}
\end{align}
This is an ordinary integral over $n$ variables $\omega_i$~($i=1,\cdots,n$). Furthermore, since $\la \chi(\mb x) \chi(\mb y)\ra=f(r_{xy})$ depends only on the distance $r_{xy}\equiv|\mb x-\mb y|$, the correlator $\la\zeta^n\ra$ is a function of distances $r_{ij}\equiv|\mb x_i-\mb x_j|$ only. In particular, this dependence encodes information of $\wt\zeta_{\omega}$ which is equivalent to $\zeta(\chi)$. Therefore, the distance dependence in the $n$-point correlator encodes information of preheating that is packed into $\zeta(\chi)$ in the modulated partial preheating scenario.

\subsection{Correlators in Modulated Partial Preheating}
\label{sec_ng_mpp}
\begin{figure}[h!]
\centering
\includegraphics[width=0.5\textwidth]{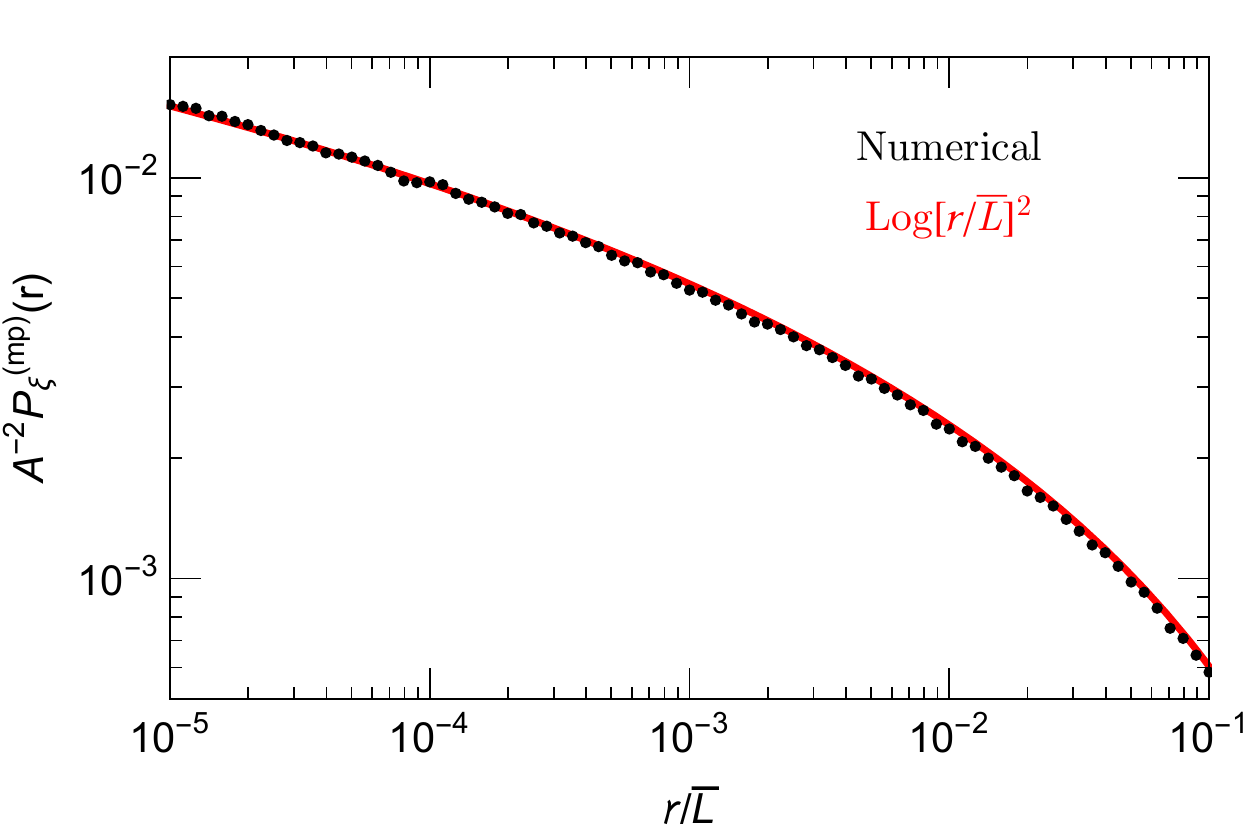} 
\caption{Normalized 2-point function $P_\zeta^\text{(mp)}(r)$ as a function of $r/{\ob L}$. We fix $\ell/L = 10^{-14}$ and $\Delta_\chi/\chi_c=0.2$. The black dots are numerical results obtained from Eq.~\eqref{eq:twopt}. Red line is the best fit assuming $P_\zeta^\text{(mp)}(r) = c_2 \log(r/L)^2$. }
\label{fig:twoptrL}
\end{figure}

\begin{figure}[h!]
\centering
\includegraphics[width=0.46\textwidth]{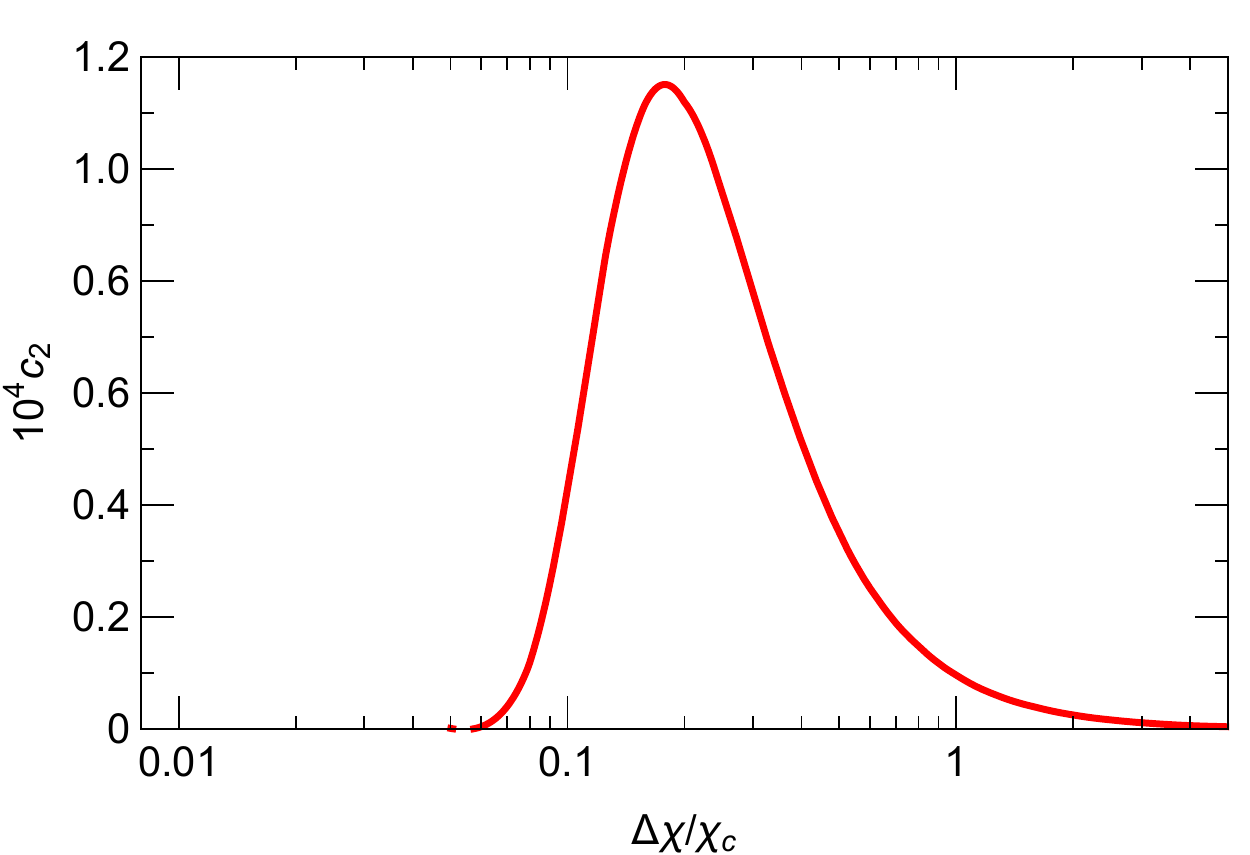} \quad \includegraphics[width=0.44\textwidth]{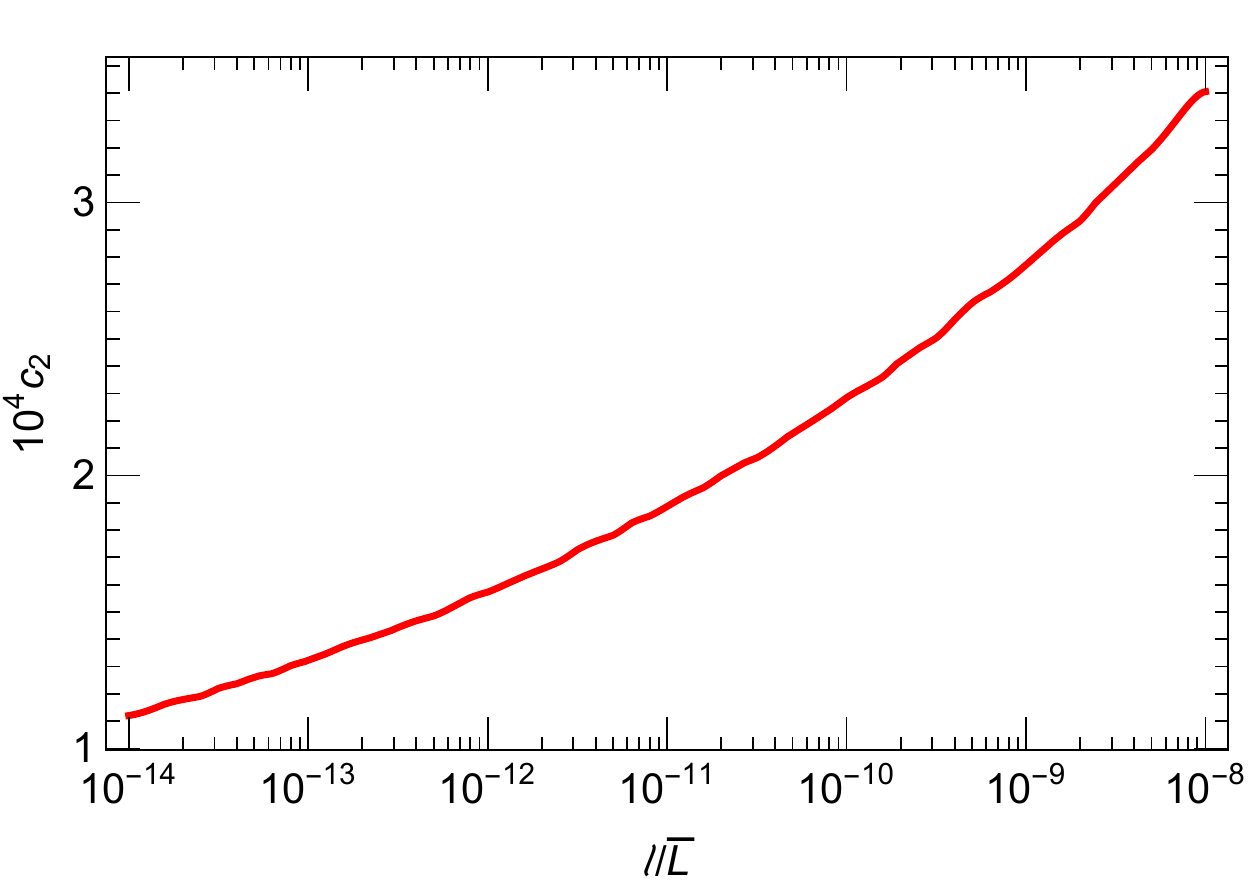} 
\caption{Left: $c_2$ as a function of $\Delta_\chi/\chi_c$, fixing $\ell/L = 10^{-14}$. Right: $c_2$ as a function of $\ell/{\ob L}$, fixing $\Delta_\chi/\chi_c = 0.2$.}
\label{fig:c2}
\end{figure}

Now we apply the formalism above to the modulated partial preheating scenario and calculate the two- and 3-point correlators.
The starting point is the function $\zeta_\text{mp}=\zeta_\text{mp}(\chi)$ in Eq.~\eqref{zetadN}. Note again that we will drop the subscript for simplicity. The Fourier transformed curvature perturbation $\wt\zeta_\omega$, defined in Eq.~\eqref{zetatilde}, is 
\bge
  \wt\zeta_\omega= 2A\FR{\sin (\chi_c \, \omega)}{\omega}.
\ede
It is more appropriate to define the perturbation with the one-point function $\bar \zeta=\la\zeta(\mb x)\ra$ subtracted, $\de\zeta=\zeta-\bar\zeta$. We then find the 2-point function, using Eq.~\eqref{chi2pt},~\eqref{eq:coincidence} and~\eqref{eq:generalformalism}, 
\begin{align}
  P_\zeta^\text{(mp)}(r)\equiv&~\la\de\zeta(\mb x)\de\zeta(\mb y)\ra=\la\zeta (\mb x)\zeta (\mb y)\ra-\la\zeta(\mb x)\ra^2\n\\
  =&\int\FR{\di\omega_1}{2\pi}\FR{\di\omega_2}{2\pi}\wt\zeta_{\omega_1}\wt\zeta_{\omega_2}\bigg(\FR{\ell}{\ob L}\bigg)^{(\omega_1^2+\omega_2^2)\Delta_\chi^2/2}\bigg[\bigg(\FR{r}{\ob L}\bigg)^{\omega_1\omega_2\Delta_\chi^2}-1\bigg],
\label{eq:twopt}
\end{align}
where $r\equiv|\mb x-\mb y|$. Note that we have used the 2-point function of $\chi$ in Eq.\eqref{chi2pt}, which is only valid when $r\ll L$. It is difficult to carry out the integral analytically and we have evaluated it numerically. By fitting to the numerical results, we find that 
\beq
P_{\zeta}^\text{(mp)}(r) \simeq A^2 c_2\bigg(\log\FR{\ell}{\ob L}, \FR{\Delta_\chi}{\chi_c}\bigg) \log^2\left(\frac{r}{\ob L}\right),
\eeq
where the coefficient $c_2>0$ is a function of $\log(\ell/\ob L)$ and $\Delta_\chi/\chi_c$. This is shown in Fig.~\ref{fig:twoptrL} and~\ref{fig:c2}. Note that the maximum value of $c_2$ is of order $10^{-4}$, which is purely a numerical fact and has no parametric dependences. 

\begin{figure}[h!]
\centering
\includegraphics[width=0.22\textwidth]{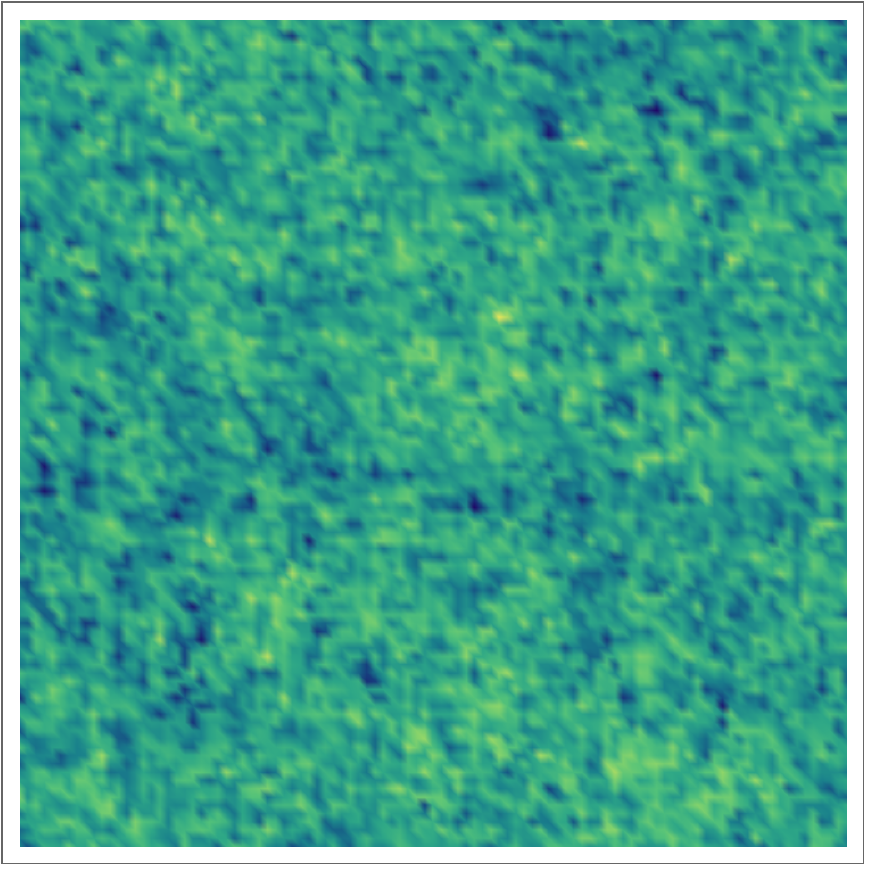}
\includegraphics[width=0.22\textwidth]{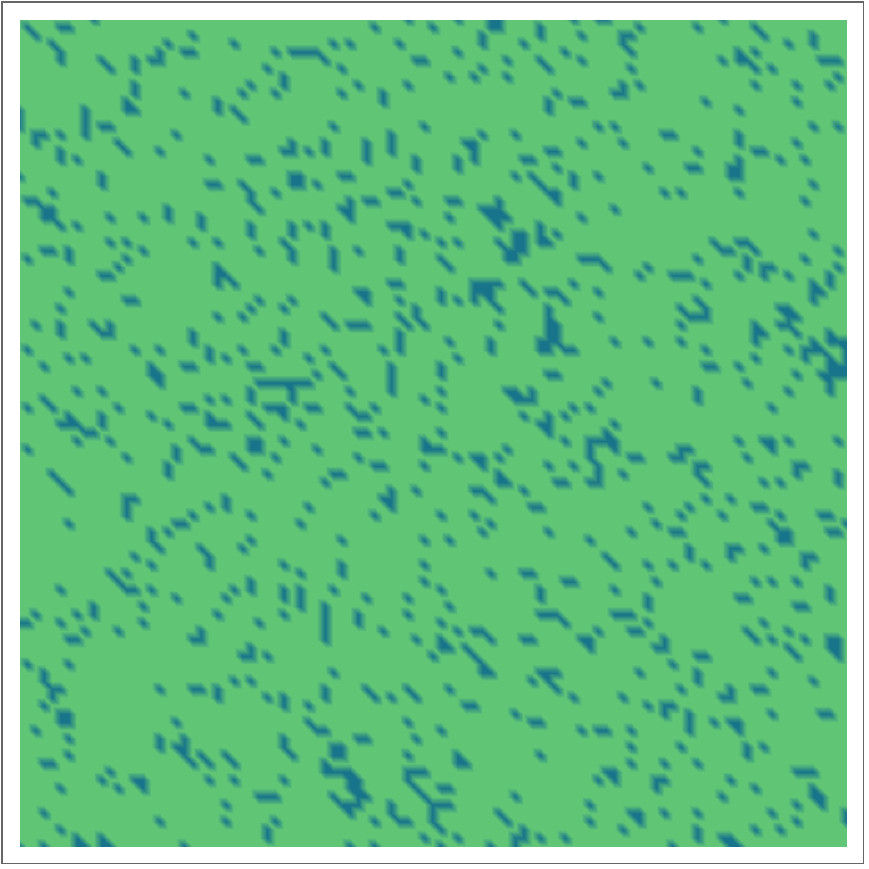}
\includegraphics[width=0.22\textwidth]{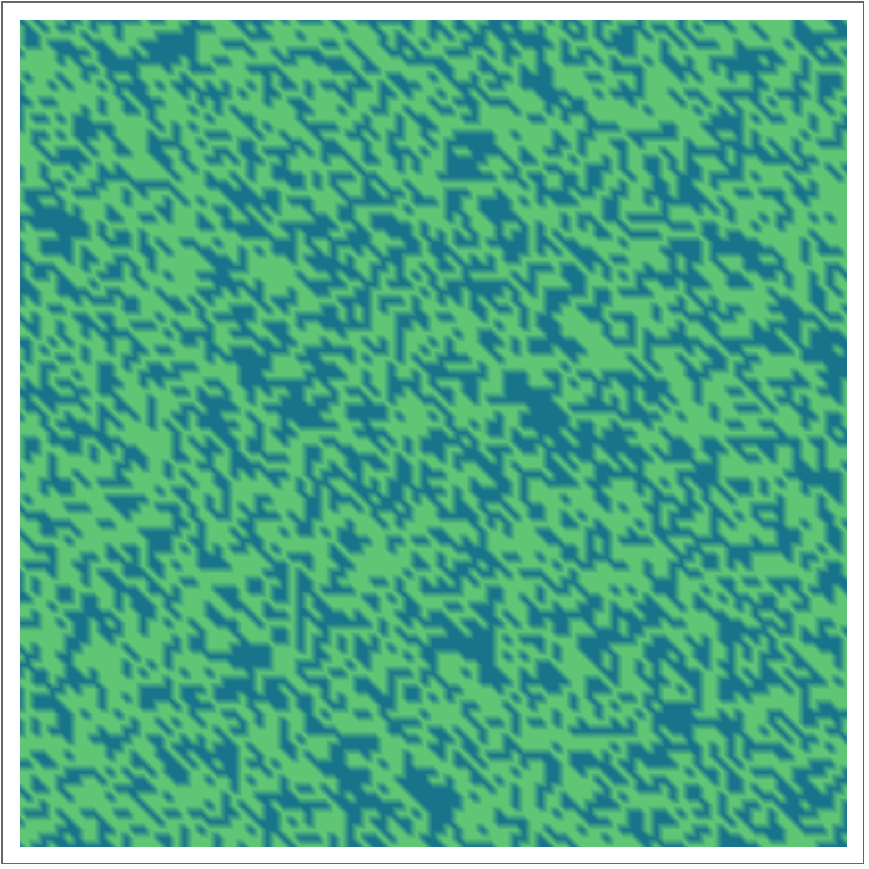}
\includegraphics[width=0.22\textwidth]{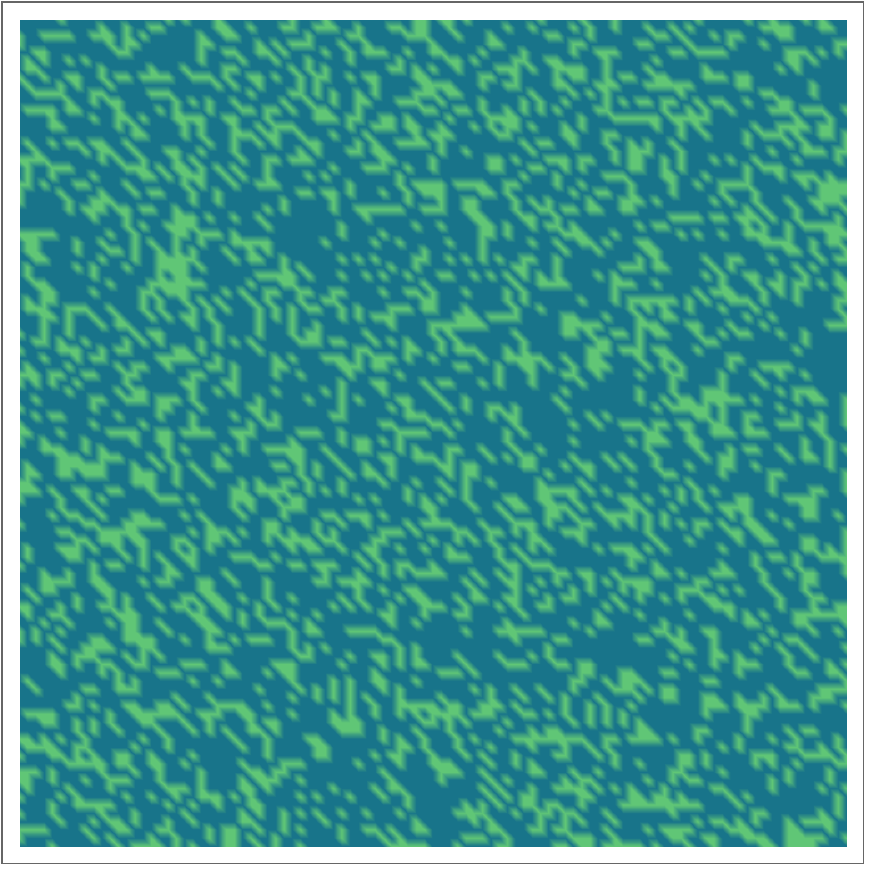}
\caption{A sample two-dimensional slice of $\chi(\mb x)$ (left) in the position space, and the corresponding $\zeta(\chi)$ for $\Delta_\chi/\chi_c=2,1,1/2$ (the right three panels). The length unit is arbitrary.}
\label{fig_fluc}
\end{figure}

We emphasize that $c_2$, and thus $P_\zeta^\text{(mp)}(r)$, depend on $\chi_c$ and $\Delta_\chi$ only through the ratio $\Delta_\chi/\chi_c$. This is also true for $n$-point functions in general. Intuitively the perturbation $\de\zeta$ should go to zero when $\Delta_\chi /\chi_c$ goes to either zero or infinity, since in these two limits the fluctuation cannot see the rectangular pulse. We demonstrate $c_2$'s dependence on $\Delta_\chi /\chi_c$ in left panel of Fig.~\ref{fig:c2}. We also show the 2D spatial slices of $P(r)$ for different choices of $\Delta_\chi/\chi_c$ in Fig.~\ref{fig_fluc}. In the right panel of Fig.~\ref{fig:c2}, we show $c_2$ as a function of $\ell/{\ob L}$. When we vary $\ell/{\ob L}$ by several orders of magnitude, $c_2$ remains the same order of magnitude. This proves that $c_2$ only depends logarithmically on $\ell/{\ob L}$. 

In momentum space, the Fourier transform of $P_\chi(r)$ gives rise to 
\beq
P_{\zeta}^\text{(mp)}(k) = \int\di^3\mb x \, e^{-\ii\mb k\cdot\mb x} P_\zeta^\text{(mp)}(r) \simeq 4\pi^2 A^2 c_2 \, \frac{\log(k \ob L)}{k^3} + \cdots
\label{eq:momentumspace}
\eeq
where we ignore unphysical oscillation terms due to the cutoff $L$ and higher order terms of ${\cal O}(k^{-4})$. This contribution has a mild scale dependence and leads to a blue-tilted spectrum. We should combine it with the inflaton contribution $P_\zeta^\text{(inf)}(k)=\la\zeta_\text{inf}(\mb k)\zeta_\text{inf}(-\mb k)\ra$, so that the total power spectrum is,
\begin{align}
\label{Ptotal}
  P_\zeta = P_\zeta^\text{(inf)} + P_\zeta^\text{(mp)}.
\end{align}
We compare Eq.~\eqref{Ptotal} with the standard parameterization of the 2-point function in the $|n_s-1| \ll 1$ limit,
\beq
P_{\zeta}(k) \simeq \frac{2 \pi^2 A_s}{k^3} \left[1+ (n_s -1) \log \left(\frac{k}{k_0}\right) \right].
\label{eq:observation}
\eeq
According to the Planck 2018 analysis~\cite{Aghanim:2018eyx}, the observed total power spectrum $P_\zeta^\text{(obs)}$ has $ A_s^\text{(obs)}\simeq 2.1\times 10^{-9}$ and $n_s^\text{(obs)} = 0.9665 \pm 0.0038$, with the pivot scale chosen at $k_0 = 0.05$Mpc$^{-1}$. On the other hand, the power spectrum from modulated reheating $P_\zeta^\text{(mp)}(k)$ gives $A_s^\text{(mp)}\simeq 2c_2A^2$ and $n_s^\text{(mp)}-1\simeq \order{1} $. The $\order{1}$ factor here comes from the difference between the chosen map size $\ob L$ and the pivot scale, which are usually quite close to each other. This ambiguity can in principle be fixed by a more careful analysis. 
In our model, 
\beq
A_s^\text{(obs)} &=& A_s^\text{(inf)} + A_s^\text{(mp)}, \nonumber \\
n_s^\text{(obs)}-1 &= & \frac{A_s^\text{(inf)}}{A_s^\text{(obs)}}\Big[ n_s^\text{(inf)}-1\Big] +  \frac{A_s^\text{(mp)}}{A_s^\text{(obs)}} \Big[n_s^\text{(mp)}-1\Big]. 
\eeq
Requiring the $A_s^\text{(obs)}$ and $ n_s^\text{(obs)}$ being consistent with observation, we get the following two constraints:
\begin{align}
2  A^2 c_2   \lesssim & ~ A_s^\text{(obs)}, \nonumber    \\
2 A^2 c_2  \lesssim & ~\left|n_s^\text{(obs)}-1\right| A_s^\text{(obs)} = 0.034 A_s^\text{(obs)},
\label{eq:twoconstraint}
\end{align}
where we drop the $\order{1}$ factor. It is clear that the second constraint is always stronger. Thus in our model, $A_s^\text{(mp)} \ll A_s^\text{(inf)} \simeq A_s^\text{(obs)}$ and $|  n_s^\text{(inf)}-1| \gtrsim | n_s^\text{(obs)}-1|$. 

\begin{figure}[h!]
\centering
\includegraphics[width=0.55\textwidth]{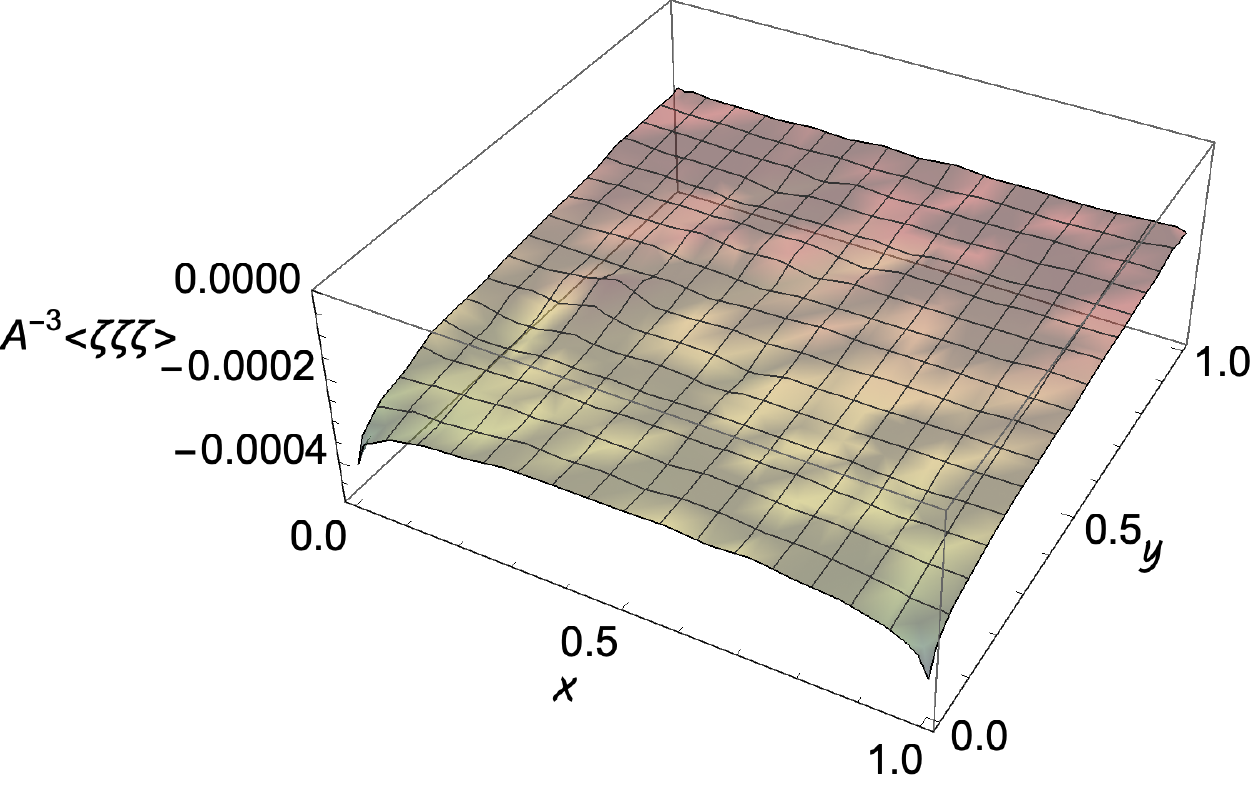}
\caption{$A^{-3} \la\de\zeta(\mb x_1)\de\zeta(\mb x_2)\de\zeta(\mb x_3)\ra$ fixing $\mb x_1 = (0,0)$, $\mb x_2 = (1,0)$ (in arbitrary length unit) and varying $\mb x_3$ in a square. }
\label{fig:3ptshape}
\end{figure}

Next we consider the 3-point function. Using the general formalism in Eq.~\eqref{eq:generalformalism}, we have 
\begin{align}
  \la\de\zeta(\mb x_1)\de\zeta(\mb x_2)\de\zeta(\mb x_3)\ra
  =&~\la\zeta(\mb x_1)\zeta(\mb x_2)\zeta(\mb x_3)\ra-\bar\zeta\Big(\la\zeta(\mb x_1)\zeta(\mb x_2)\ra+2\;\text{perms}\Big)+2\bar\zeta^3\n\\
  =&\int\prod_{i=1}^3\bigg[\FR{\di\omega_i}{2\pi}\wt\zeta_{\omega_i}\bigg(\FR{\ell}{L}\bigg)^{\omega_i^2\Delta_\chi^2/2}\bigg]\Big(C_{12}C_{23}C_{31}-C_{12}-C_{23}-C_{31}+2\Big),
\end{align}
where
\bge
  C_{ij}\equiv\bigg(\FR{r_{ij}}{\ob L}\bigg)^{\omega_i\omega_j\Delta_\chi^2},~~~r_{ij}\equiv|\mb x_i-\mb x_j|.
\ede
Numerically it turns out that the 3-point function does not depend much on the shape of the triangle connecting the three points. It is shown in Fig.~\ref{fig:3ptshape} that the values of $A^{-3} \la\de\zeta(\mb x_1)\de\zeta(\mb x_2)\de\zeta(\mb x_3)\ra$ are roughly the same varying a point while fixing the positions of the other two points. 

\begin{figure}[h!]
\centering
\includegraphics[width=0.5\textwidth]{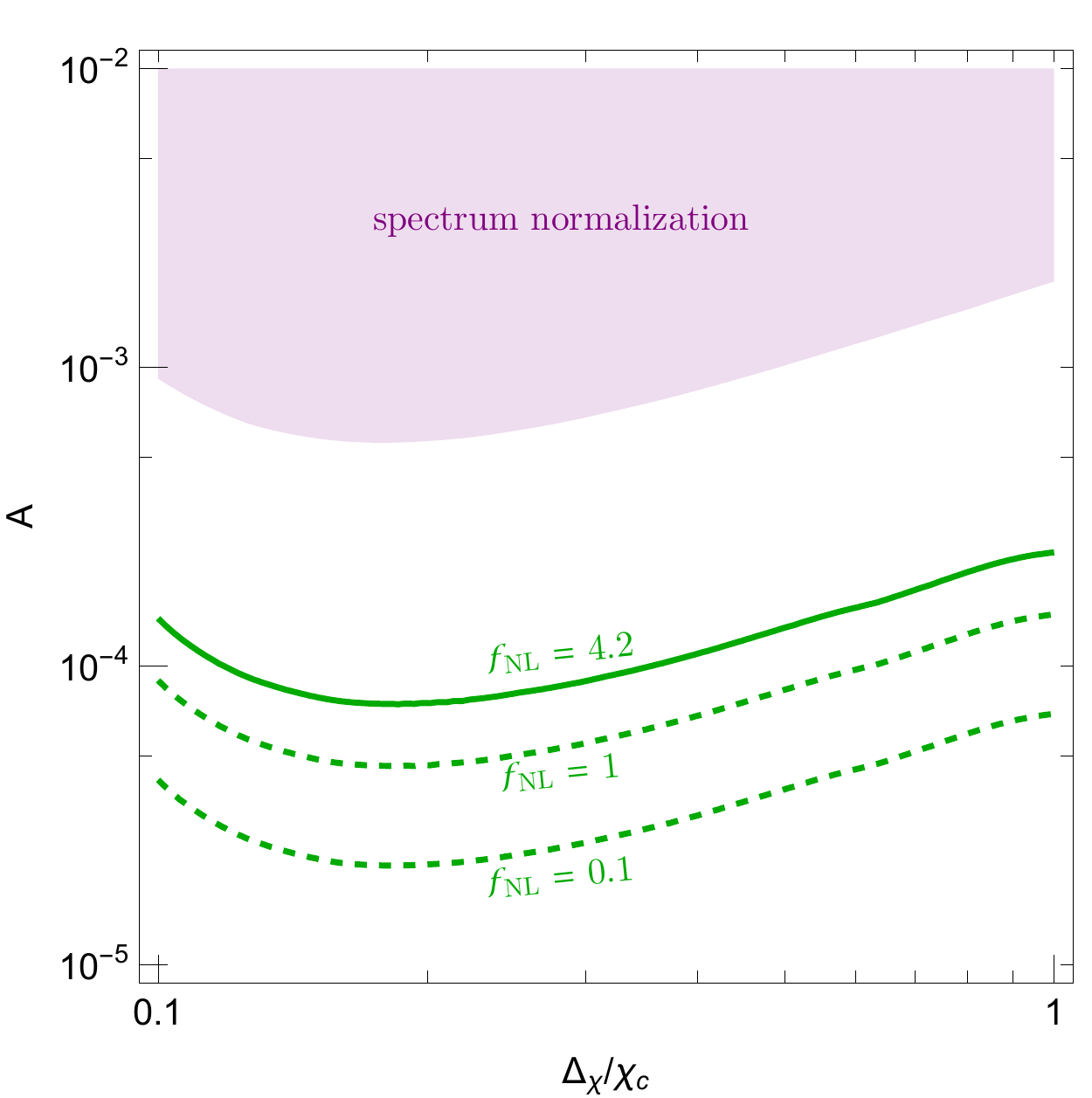}
\caption{Allowed region in the $(\Delta_\chi/\chi_c, A)$ plane. The purple region is excluded by the spectrum normalization constraints in Eq.~\eqref{eq:twoconstraint}. The green dashed lines show the predicted $f_{NL}$ based on Eq.~\eqref{eq:fNL}. The green solid line corresponds to the current $1\sigma$ upper bound on local non-Gaussianity~\cite{Akrami:2019izv}, above which the parameter space is excluded. We fix $\ell/L = 10^{-14}$, $\log(k_3^{(0)} \ob L)$ = 20, and $\log(k_1^{(0)} \ob L)=10$. The results will not be affected much by other choices of these parameters. }
\label{fig:para}
\end{figure}

It is useful to consider the squeezed limit of the 3-point function with $r_\ell \equiv r_{12} = r_{13} \gg r_{23} \equiv r_s$. Numerically we find that in this limit, we have
\beq
 \la\de\zeta(\mb x_1)\de\zeta(\mb x_2)\de\zeta(\mb x_3)\ra \simeq A^3 \left[c_{31} \log\left(\frac{r_s}{\ob L}\right) + c_{32} \log^2\left(\frac{r_s}{\ob L}\right)\right] \log^2\left(\frac{r_\ell}{\ob L}\right),
\eeq
where $c_{31}$ and $c_{32}$, analogous to $c_2$, are functions of $\log (\ell/\ob L)$ and $\Delta_\chi/\chi_c$. The maximum values of $c_{31}$ and $c_{32}$ are also suppressed numerically and are of order $10^{-6}$. 
In momentum space, this corresponds to 
\beq
\lim_{k_1/k_3 \to 0} \la\de\zeta(\mb k_1)\de\zeta(\mb k_2)\de\zeta(\mb k_3)\ra \simeq (2 \pi)^7 \delta^3(\mb k_1 + \mb k_2 + \mb k_3) A^3 \frac{ \left[ c_{31}+ 2 c_{32} \log(k_3 \ob L)  \right] \log(k_1 \ob L) }{2k_1^3 k_3^3}. 
\eeq
This is close to the form of the local non-Gaussianity~\cite{Komatsu:2001rj}. Ignoring the mild scale dependence, we could estimate the local $f_\text{NL}$ to be 
\beq
f_\text{NL} = \frac{5A^3}{6A_s^2}  \left[ c_{31}+ 2 c_{32} \log(k_3^{(0)} \ob L)  \right] \log(k_1^{(0)} \ob L), 
\label{eq:fNL}
\eeq
where we pick some fixed $k_3^{(0)}$ and $k_1^{(0)}$. Combining Eq.~\eqref{eq:twoconstraint} and Eq.~\eqref{eq:fNL}, we show the allowed parameter space in Fig.~\ref{fig:para}. One could see that even when $A  \simeq  \gamma/3 \left(\Gamma_\phi/\Gamma_\si \right)^{2/3} \sim {\cal O} (10^{-4})$, the predicted non-Gaussianity is $\sim {\cal O}(1)$, within the reach of next generation cosmological measurements~\cite{Abazajian:2019eic}. 

\section{Conclusion and Outlook}
\label{sec_disc}

In this paper we have studied a scenario of modulated partial preheating, in which the space variation of a modulating field $\chi$ triggers the preheating of a spectator field $\si$ after inflation. This scenario generates characteristic local non-Gaussianities at large scales with observably large $f_\text{NL}$. This provides a unique chance to observe the nonlinear dynamics of preheating era which is in general difficult to probe directly. We have showed that the expansion histories of local patches during preheating have a two-phase structure controlled by the background value of the modulating field $\chi$: the patches with $|\chi|$ larger than a critical value $\chi_c$ have efficient particle production, while patches with $|\chi|<\chi_c$ do not. This induces ``square wave'' behavior in the curvature perturbation, which is a unique feature of the non-perturbative dynamics. We have also calculated the corresponding power spectrum as well as non-Gaussianity of this ``square wave'' component, and found it to be nearly scale invariant and with fairly sizable local non-Gaussianity. Another potential observable of this scenario is the isocurvature mode of the modulating field $\chi$, if it remains today as a fraction of dark matter. But this is very model dependent. For instance, the isocurvature mode of $\chi$ could disappear completely if $\chi$ decays after reheating of the inflaton.

One may want to search for this effect by distilling the bivalued square wave distribution directly out of the observed fluctuations, as shown in Fig.\;\ref{fig_fluc}. But this is likely impossible. The reason is that the bivalued distribution depends on the ``UV cutoff'' $\ell$. Physically, this UV cutoff is provided by the horizon size of each Hubble patch during preheating, which is way below any resolution we can imagine for observing large-scale fluctuations. Consequently, what we actually observe is necessarily the average of the bivalued distribution over many Hubble patches at the time of preheating. The averaged distribution is no longer bivalued. So we have to look for effects that survives this average, among which the local non-Gaussianity is a good example. It survives the average because it has very weak scale dependence and is present practically at all scales. It is interesting to explore other possible observables of this effects that survive the horizon averaging.

More open questions and directions remain to be explored:
\begin{itemize}
\item Couplings in low-energy effective theories, instead of being constants, could originate from dynamical fields in their UV completions. The time evolution of such fields could lead to quite a variety of novel phenomenon, which has always been an interesting direction of model building in the particle physics community. What we have explored instead is to use the {\it spatial} variation of a dynamical field governing a coupling to probe the phase diagram of nonlinear dynamics in the early universe. Could our toy model be embedded in a full-fledged particle physics model? What will happen if the modulating field in our model is the SM Higgs? 
\item Recently the use of non-Gaussianity to probe particle physics, in particular heavy particles beyond the reach of ordinary colliders, has been developed rapidly in the context of cosmological collider physics~\cite{Chen:2009zp,Baumann:2011nk,Noumi:2012vr,Arkani-Hamed:2015bza,Chen:2016nrs,Lee:2016vti,Meerburg:2016zdz,Chen:2016uwp,Chen:2016hrz,An:2017hlx,Kumar:2017ecc,Chen:2018xck,Wu:2018lmx,Li:2019ves,Lu:2019tjj,Hook:2019zxa,Hook:2019vcn,Kumar:2019ebj,Wang:2019gbi,Wang:2020uic,Li:2020xwr,Wang:2020ioa}. What we have studied in this paper is, to a certain extent, a {\it non-perturbative} version of the cosmological collider. What other phase diagrams of non-perturbative processes in the early universe could be probed using our modulating mechanism? We have only studied the 3-point function in the squeezed limit. How about non-Gaussianity of general shapes and higher point correlations? 
\end{itemize}

\paragraph{Acknowledgements}
We thank Mustafa Amin, Kaloian Lozanov, Matt Reece, and Yi Wang for useful discussions and comments.   
JF is supported by the DOE grant DE-SC-0010010 and NASA grant 80NSSC18K1010.  

\appendix 

\section{Energy Transfer in the Linearized Regime}
\label{app:fraction}

The energy density transferred from $\si$ to $\psi$ in the co-moving volume is given by 
\beq
\rho_\psi(t) a(t)^3 = \int \frac{d^3 k}{(2\pi)^3} \left|\omega_k \right| n_k,
\label{eq:rhoestimate}
\eeq
where the effective frequency is given by 
\beq
\omega_k^2(t) = \frac{k^2}{a^2} + \frac{g \sigma_0} {a^{3/2}}\cos\left[m_\sigma(t-t_0)\right] = \frac{k^2}{\ti^{4/3}}+ \frac{q_0 m_\si^2}{2 \ti}\cos\left[2(\ti-1)\right]. 
\eeq
Note that the effective mass term, $m_{\rm {eff}}^2 =\frac{q_0 m_\si^2}{2 \ti}\cos\left[2(\ti-1)\right]$ oscillates. Ignoring the Hubble expansion in a single cycle, the cycle average of the effective mass, $\overline{m_{\rm {eff}}^2}$ is approximately zero. Thus averaging over a cycle,
\beq
\overline{\omega_k^2(t)} \simeq \frac{k^2}{\ti^{4/3}}. 
\label{eq:omega}
\eeq
Consequently, we take $\psi$ particles to be relativistic. 
Combining Eq.~\eqref{eq:rhoestimate} and Eq.~\eqref{eq:lnnk}, \eqref{eq:kmax}, \eqref{eq:omega}, we have the fraction of energy in $\psi$ as 
\beq
\rho_\psi(t) a(t)^3 &=& e^{\frac{8 \alpha}{\pi} \sqrt{q_0 \ti}} \int_0^{k_{\rm max}} \frac{dk}{2 \pi^2} \frac{k^3}{\ti^{\, 2/3}} e^{-\frac{48 \alpha \, k^2}{\pi \sqrt{q_0} m_\si^2} \ti^{\; 1/6}}, \nonumber \\
&\simeq & e^{\frac{8 \alpha}{\pi} \sqrt{q_0 \ti}} \frac{q_0 m_\si^4} {9216 \alpha^2 \ti}, \nonumber \\
\delta(t) &\equiv & \frac{\rho_\psi(t)}{\rho_{\si+\psi} }= \frac{2\rho_\psi(t) a(t)^3}{m_\si^2 \si_0^2} = e^{\frac{8 \alpha}{\pi} \sqrt{q_0 \ti}} \frac{q_0} {4608 \alpha^2 \ti} \left(\frac{m_\si}{\si_0}\right)^2, 
\eeq
where $\alpha \simeq 0.85$ and $\ti = m_\si t/2$. Note that the estimate above is only valid for $q_0 \gtrsim 5$. For smaller $q_0$, there is negligible particle production due to the expansion of the Universe and Eq.~\eqref{eq:lnnk} doesn't apply. 

\begin{figure}[h]
\centering
\includegraphics[width=0.5\textwidth]{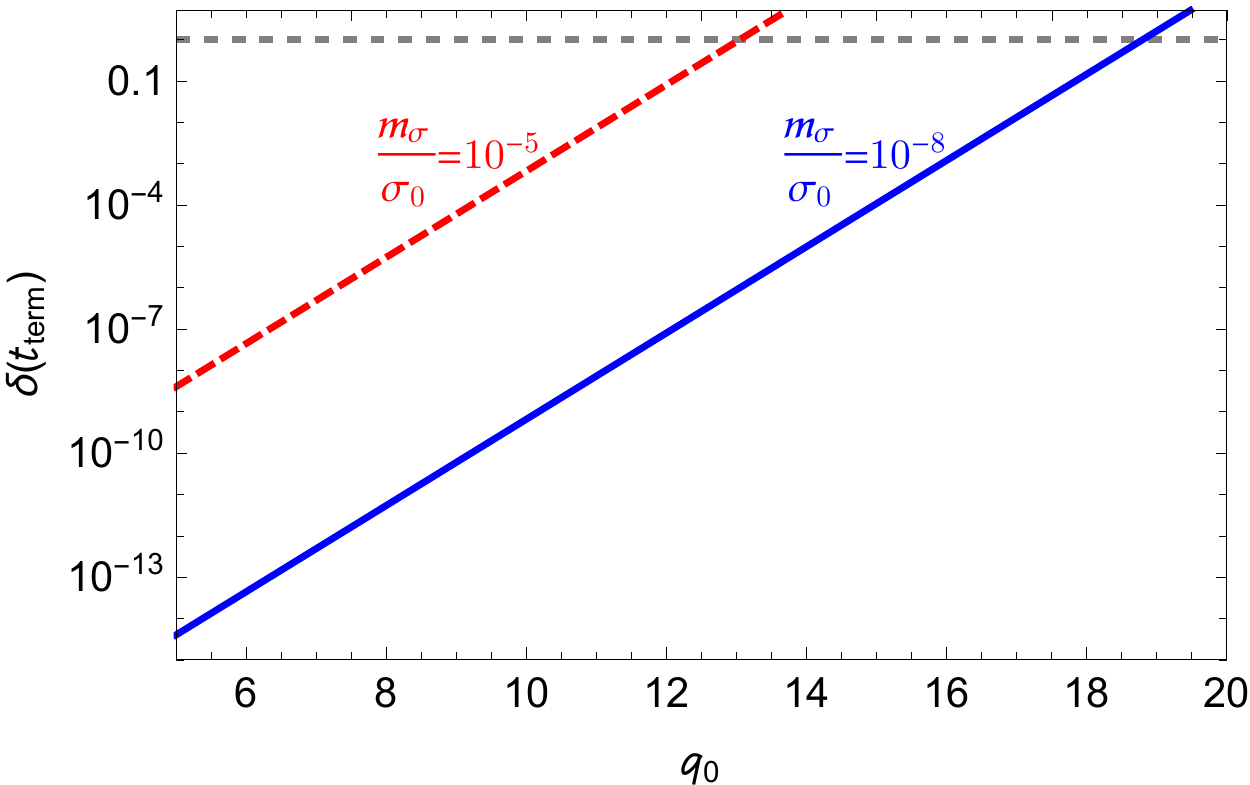} 
\caption{$\rho_\psi/\rho_{\si+\psi}$ as a function of $q_0$ for three different values of $m_\si/\si_0$ based on the estimate in Eq.~\eqref{eq:delta}. The grey dashed horizontal line indicates 1. When $\rho_\psi/\rho_{\si+\psi}$ gets close to the line, feedback from $\psi$ to $\si$ is effective.}
\label{fig:fraction}
\end{figure}

\section{Lattice Simulations With a Fixed Cosmic Expansion}
\label{app:lattice}
\begin{figure}[h]
\centering
\includegraphics[width=0.5\textwidth]{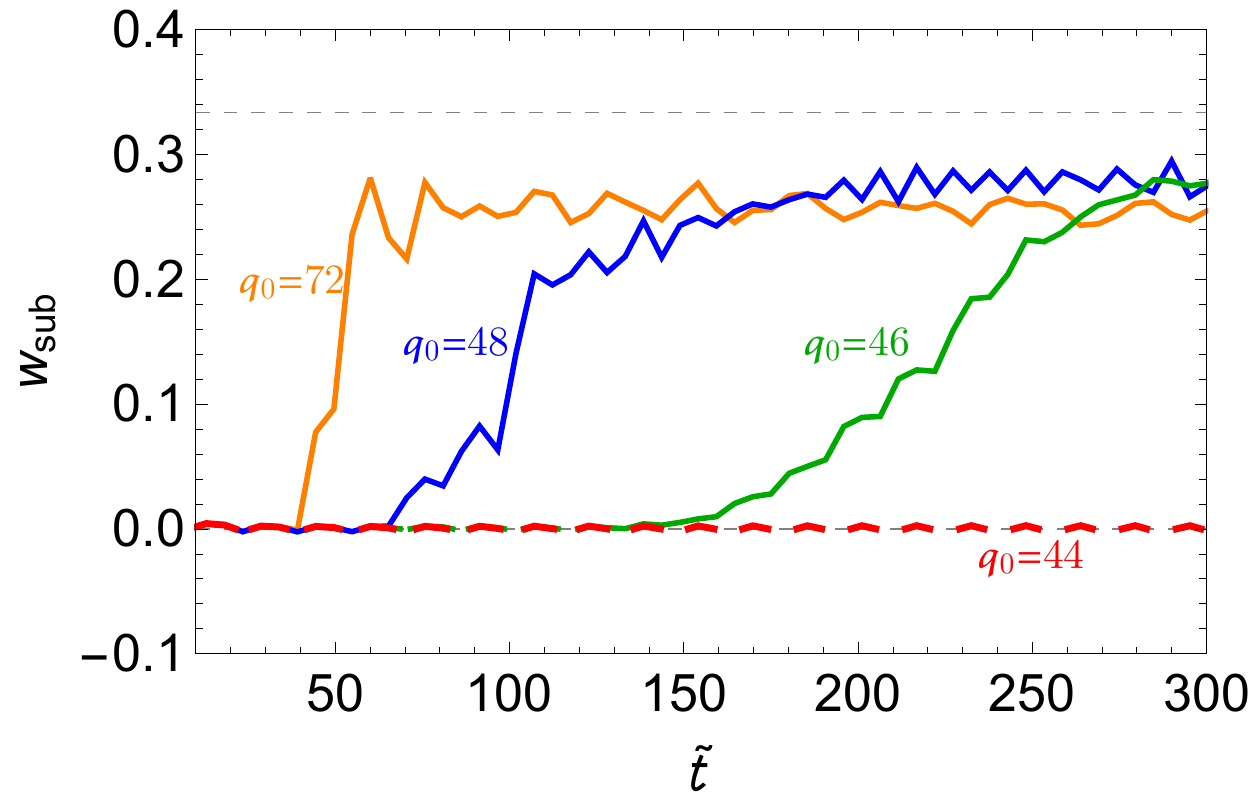} 
\caption{Effective equation of state for different $q_0$'s in the sub-system $\si$ and $\psi$. We fix $m_\si/\si_0 = 10^{-13}$ and $\lambda = 6.5 \times 10^{-24}$. }
\label{fig:latticeresults}
\end{figure}

We use LatticeEasy~\cite{Felder_2008} to simulate the evolution of the spectator field $\si$ and its daughter field $\psi$ in a matter-dominated background. The results of the effective equation of state $w_{\rm {sub}} = p_{\rm {sub}}/\rho_{\rm {sub}}$ for the sub-system are shown in Fig.~\ref{fig:latticeresults}. Indeed for the given parameters, varying $q_0$ (or equivalently, $g$) a bit around $q_c \sim 46$, the effective equation of state varies from close to $1/3$ to 0. 

\providecommand{\href}[2]{#2}\begingroup\raggedright\endgroup

\end{document}